\documentclass[12pt,a4paper]{article}
\usepackage{graphicx,amssymb}
\newcommand{\be}{\begin{equation}}
\newcommand{\ee}{\end{equation}}
\newcommand{\bea}{\begin{eqnarray}}
\newcommand{\eea}{\end{eqnarray}}

\catcode`@=12
\def\la{\mathrel{\mathchoice {\vcenter{\offinterlineskip\halign{\hfil
$\displaystyle##$\hfil\cr<\cr\sim\cr}}}
{\vcenter{\offinterlineskip\halign{\hfil$\textstyle##$\hfil\cr<\cr\sim\cr}}}
{\vcenter{\offinterlineskip\halign{\hfil$\scriptstyle##$\hfil\cr<\cr\sim\cr}}}
{\vcenter{\offinterlineskip\halign{\hfil$\scriptscriptstyle##$\hfil\cr<\cr\sim
\cr}}}}}

\begin {document}
\begin {flushright}
SINP-APC-11/01
\end {flushright}
\thispagestyle {empty}
\begin {center}
{\Large\bf 
{The Real Gauge Singlet Scalar Extension of Standard Model: A
Possible Candidate of Cold Dark Matter}}\\
\vspace{1cm}
{{\bf Anirban Biswas} \footnote{email: anirban.biswas@saha.ac.in}, 
{\bf Debasish Majumdar} \footnote{email: debasish.majumdar@saha.ac.in}}\\
\vspace{0.25cm}
{\normalsize \it Astroparticle Physics and Cosmology Division,}\\
{\normalsize \it Saha Institute of Nuclear Physics,} \\
{\normalsize \it 1/AF Bidhannagar, Kolkata 700064, India}\\
\vspace{1cm}
{\bf ABSTRACT}
\end {center}
We consider a simplest extension of Standard Model in which a real SM 
gauge singlet scalar with an additional discrete symmetry $Z_2$ is 
introduced to SM. This additional scalar can be a viable candidate of 
cold dark matter since the stability of $S$ is achieved by the application 
of $Z_2$ symmetry on $S$. Considering $S$ as a possible candidate of 
cold dark matter we have solved Boltzmann's equation to find the 
freeze out temperature and relic density of $S$ for Higgs mass 120 GeV 
in the scalar mass range 5 GeV to 1 TeV. 
As $HHSS$ coupling $\delta_2$ appearing in 
Lagrangian depends upon the value of scalar mass $m_S$ and Higgs mass 
$m_h$, we have constrained the $m_S - \delta_2$ parameter space by using 
the WMAP limit on the relic density of dark matter in the universe and 
the results of recent ongoing dark matter direct search experiments 
namely CDMS-II, CoGeNT, DAMA, EDELWEISS-II, XENON-10, XENON-100. 
From such analysis we find two distinct mass regions (a lower and 
higher mass domain) for such a dark matter candidate that satisfy both the 
WMAP limit and the experimental results considered here.  
We have estimated the possible differential direct detection 
rates and annual variation of total detection rates for 
this scalar dark matter candidate $S$ for two detector materials
namely Ge, Xe. Finally we have calculated the $\gamma-$ray flux from
the galactic centre due to annihilation of two 130 GeV scalar dark matter into
two monoenergetic $\gamma-$rays.
\vskip 1cm
\qquad\quad Pacs: 95.35.+d, 98.80.Cq
\vskip 1cm
\quad\,\, Dark Matter, Beyond SM
\newpage
\section {Introduction}
In recent years, one of the most important areas of modern cosmology is 
to investigate the existence and nature of dark matter in the universe. 
The observations 
by Wilkinson Microwave Anisotropy Probe (WMAP) \cite{wmap} for studying the 
fluctuations in cosmic microwave background radiation reveal that the 
universe consists of 27\% matter and the rest 73\% is an unknown energy
known as Dark Energy. Out of this 27\%, only about 4\% accounts for the ordinary
matter like leptons and baryons, gas, stars and galaxies etc. The rest 
about 23\% matter is completely unknown. Moreover there are several 
cosmological observations like rotation curves of spiral galaxies,
the gravitational micro-lensing, observations on Virgo \cite{virgo}
and Coma clusters \cite{coma}, bullet clusters \cite{bullet},
etc. which provide indications of the existence 
of huge amount of non-luminous matter or dark matter (DM) in the universe. 

Nature and identity of the constituents of dark matter are mostly 
unknown. However, evidences suggest that the dark matter candidates  
are mostly stable, non-baryonic, massive, non-relativistic 
particles having negligible or very weak interactions with other particles. 
These types of dark matter are often termed as cold dark matter (CDM) 
or weakly interacting massive particles (WIMP). In the early universe, 
these particles would have been present in large numbers in thermal 
equilibrium. As the universe expands and cools down their density decreases 
resulting in decrease in their interaction/annihilation rates. 
When the expansion rate of the universe becomes larger than the annihilation 
rate of the WIMPs, they get decoupled from the universe. Thus they 
``freeze out" from the other contents of the universe and remain as relics. 
The temperature at which this phenomenon occurs is known as ``freeze out" 
temperature and its density is called ``relic density". After freeze out, 
the relic density of WIMP is only affected by the expansion of the universe. 
Since Standard Model (SM) of particle physics cannot provide any viable 
candidate for cold dark matter, one has to consider theories beyond SM in 
order to explain the dark matter candidates (namely WIMP). 
 
In this paper we have considered the simplest possible renormalisable 
extension of SM by adding a real gauge singlet scalar $S$. We impose a 
discrete symmetry $Z_2$ on $S$ and due to this symmetry the additional 
scalar $S$ is stable and can be a viable candidate for cold dark matter.
This model was first proposed by V. Silveira and A. Zee \cite{zee}. 
Thereafter a number of authors have explored its phenomenology \cite{earlier}. 
The relevance of the scalar singlet as a plausible candidate for dark matter
is very elaborately described in Ref. \cite{tytgat1} (and references therein). 
Investigating the relic density of a scalar dark matter by constraining the 
unknown parameters from direct detection experiments are addressed by previous 
authors. In Ref. \cite{tytgat2}, the relic density is investigated for 
scalar singlet by constraining dark matter mass and direct detection rates 
from DAMA \cite{DAMA} results.
Similar analysis including the CoGeNT \cite{cogent10} results and 
CDMS II \cite{cdms2} results are also addressed in Ref. \cite{tytgat3}. The 
analysis of scalar singlet dark matter scenario for XENON 100 \cite{XENON100}
direct detection experiment results are also given in this reference.  
The scalar singlet dark matter with CoGeNT results are also discussed by 
Fitzpatrick et al \cite{fitzpatrick}. The interpretation of Fermi-Lat 
results \cite{fermilat} with scalar singlet dark matter is discussed 
in Ref. \cite{tytgat4}.

In the present work we estimate the freeze out temperature and relic density 
of the dark matter candidate $S$ by solving Boltzmann's equation. Then we 
constrain the parameter $\delta_2$ \footnote {only parameter in this model 
which appears in both the expressions of scattering and annihilation cross 
section of $S$ and which depends on the  masses of scalar $S$ and Higgs $h$}
by using WMAP limit on relic density of dark matter and the results of
recent dark matter direct detection experiments like CDMS-II
\cite{cdms2, cdms2-10}, XENON-10 \cite{XENON10}, XENON-100 \cite{XENON100},
CoGeNT \cite{cogent10,cogent11}, EDELWEISS-II \cite{EDELWEISS}
and DAMA \cite{DAMA}. In CDMS and CoGeNT experiments the target
material is Ge and in XENON experiments the target materials is Xe. 

The constrained parameters thus obtained are then used to calculate
the differential direct detection rates and the annual variation of
total detection rates of the scalar dark matter candidate
$S$ for two detector materials namely Ge, Xe. Therefore
we have calculated the $\gamma-$ray flux due to
130 GeV scalar dark matter for the annihilation channel $SS\rightarrow\gamma\gamma$
from the galactic centre.

The paper is organised as follows. In Section 2 we give a brief 
description of the scalar singlet model. Section 3 describes the formalism 
for computing relic abundance of a particular particle candidate. The 
results of the relic density calculations are given in Section 4. The 
model parameter $\delta_2$ is constrained using the WMAP relic density 
data and the results obtained from various dark matter direct detection 
experiments. This is described in Section 5. The formalism for the calculation 
of direct detection rates and the annual variations of these rates is 
described in Section 6. With the constrained model parameter, $\delta_2$ 
as obtained in Section 5, the direct detection rates and their annual 
variations of total detection rates are calculated for this scalar dark 
matter candidate for some reference detector materials namely Ge, Xe. 
The calculational procedure and the results are described in Section 7. 
In section 8 we have calculated the $\gamma-$ray flux from galactic
centre due to annihilation of dark matter present in the galactic halo.
Finally in Section 9, we give a summary and conclusion.    
\section{The Model}
In the  present work we consider a simplest extension of Standard Model where
a real singlet scalar is added to the scalar sector of SM and explore
the possibility that it can be  a candidate for cold dark matter. 
The most general form of the potential appearing in the Lagrangian
density for such a scalar fields is
\begin{eqnarray}
\quad\qquad V(H,S) = \frac{m^2}{2}{H^\dag}H +\frac{\lambda}{4}(H^ \dag H)^2 + 
\frac{\delta_1}{2} H^\dag HS + \frac{\delta_2}{2}H^\dag H S^2 
+\left(\frac{\delta_1 m^2}{2\lambda}\right)S 
+ \frac{k_2}{2}S^2 + \frac{k_3}{3}S^3 +\frac{k_4}{4}S^4  \nonumber \\ 
\end{eqnarray}
and the Lagrangian of this model is given by 
\begin{eqnarray}
\mathcal{L} = \mathcal{L}_{\rm SM} + \frac{1}{2}\partial_\mu S
\partial^\mu S - \frac{\delta_1}{2} H^\dag HS - 
\frac{\delta_2}{2}H^\dag H S^2 \label{1} 
-\left(\frac{\delta_1 m^2}{2\lambda}\right)S
-\frac{k_2}{2}S^2 -\frac{k_3}{3}S^3 -\frac{k_4}{4}S^4  \nonumber \\
\end{eqnarray}
Where $\mathcal{L}_{\rm SM} $ is the Standard Model (SM) Lagrangian,
$H$ is the SM Higgs doublet and $S$ is the real gauge 
(SU(2)$_{\rm L}\times$U(1)$_{\rm Y}$) singlet scalar. The stability 
of $S$ will be achieved by imposing a $Z_2$ symmetry ($S \rightarrow -S $, 
$\mathcal{L} \rightarrow  \mathcal{L}$) over $S$. Therefore, under this 
symmetry the coefficients of odd powers of $S$ are zero 
($k_3$ and $\delta_1$ in Eq. (\ref{1})). 
After spontaneous symmetry breaking 
masses of the Scalar field $S$ and physical Higgs $h$ are  
\bea
m_S^2 &=& k_2 + {\delta_2 {\rm V}^2}/{2}\,\, , 
\label{3}
\eea
\bea
m_h^2 &=& -m^2 = {\lambda {\rm V}^2}/{2}\,\, ,
\label{4}
\eea
V is the VEV of Higgs (V = 246 GeV).
In the present work we have taken the mass $m_S$ of the 
scalar particle $S$ in the range 5 GeV - 1 TeV. Depending on its mass  
the dark matter candidate $S$ annihilates into fermion pairs, gauge boson
pairs and Higgs pairs.
\section{Formalism for Calculation of Relic Abundance}
\begin{figure}[h]
\centering
\includegraphics[width=5cm,height=3cm]{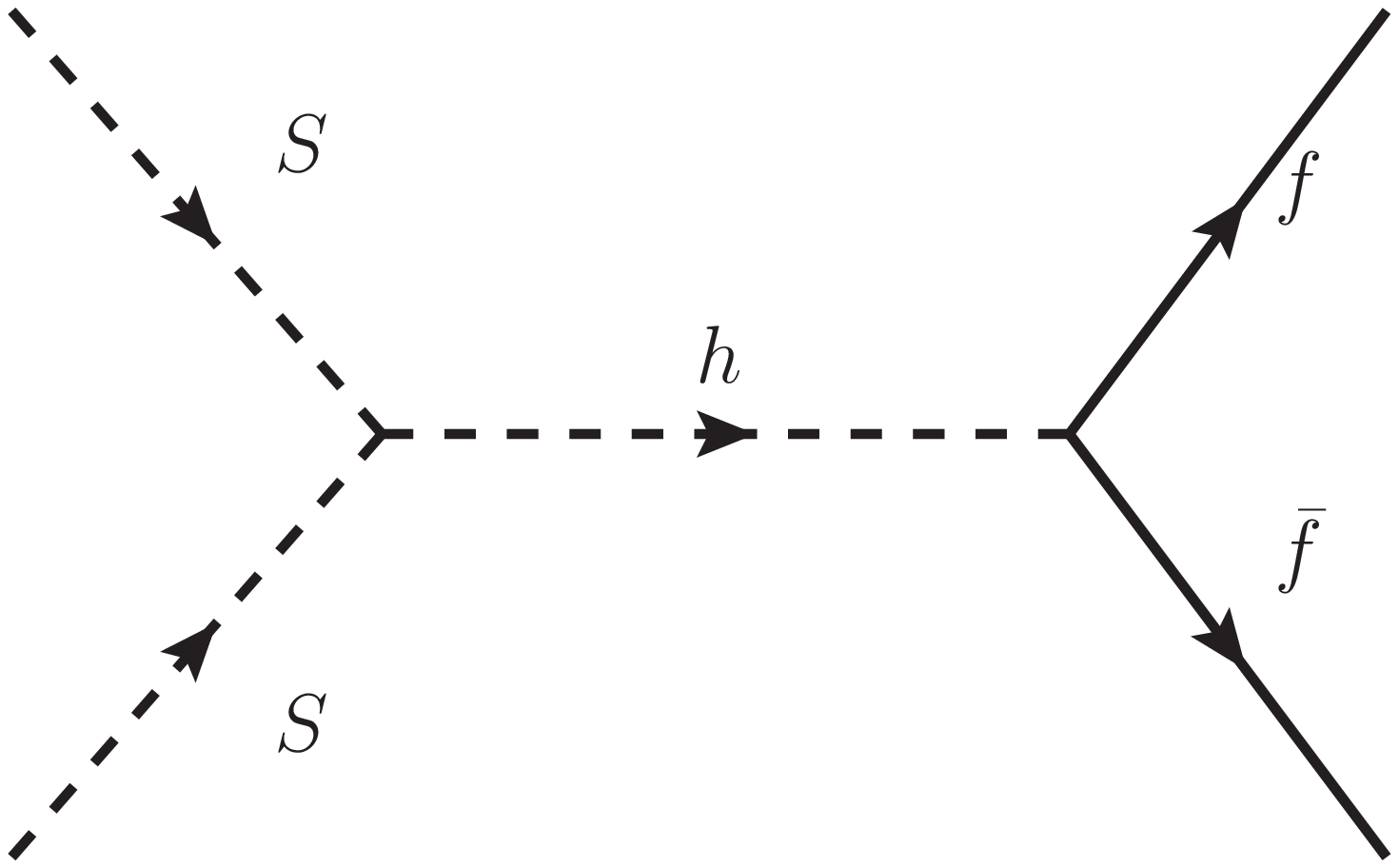}
\includegraphics[width=5cm,height=3cm]{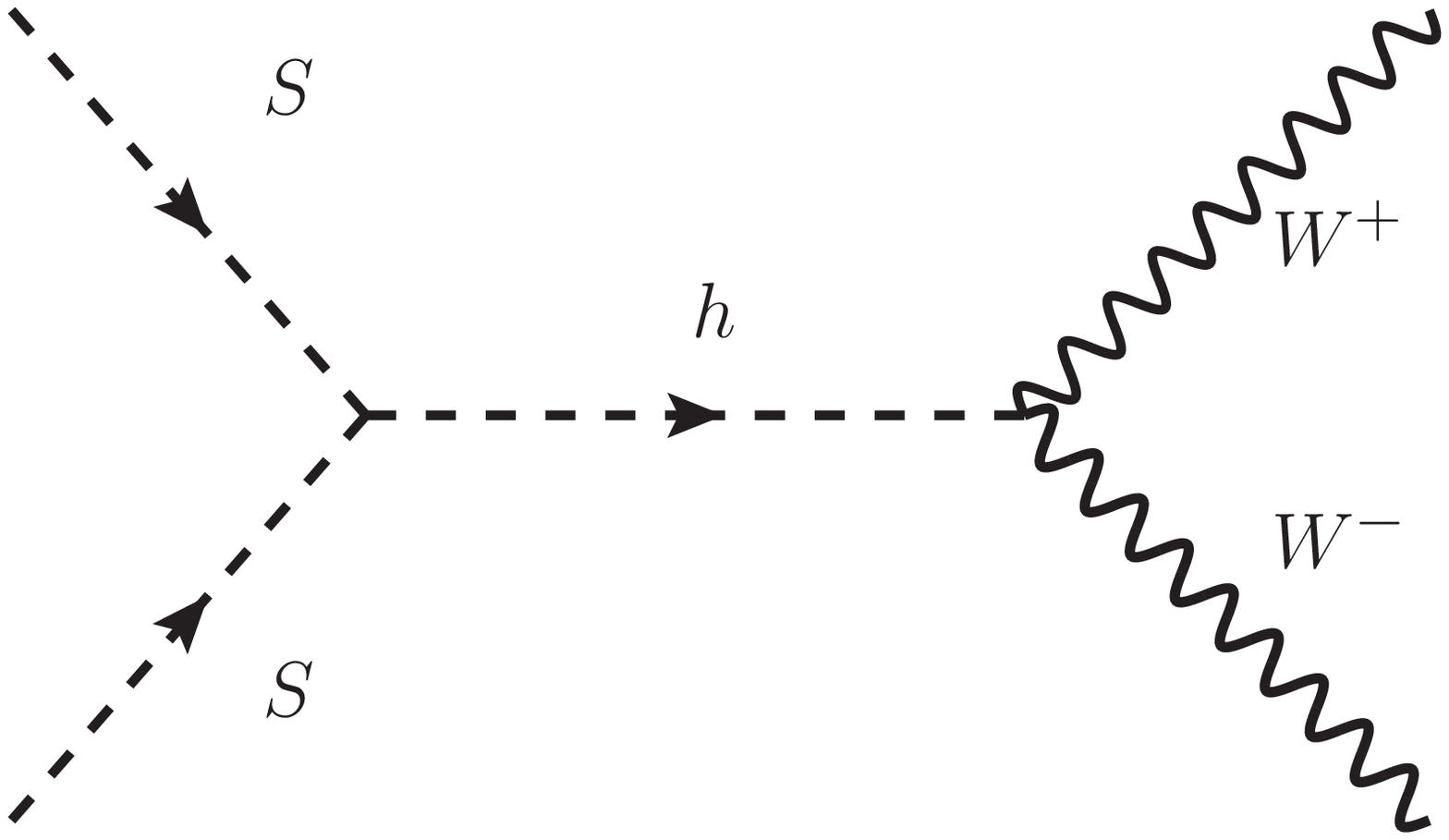}
\includegraphics[width=5cm,height=3cm]{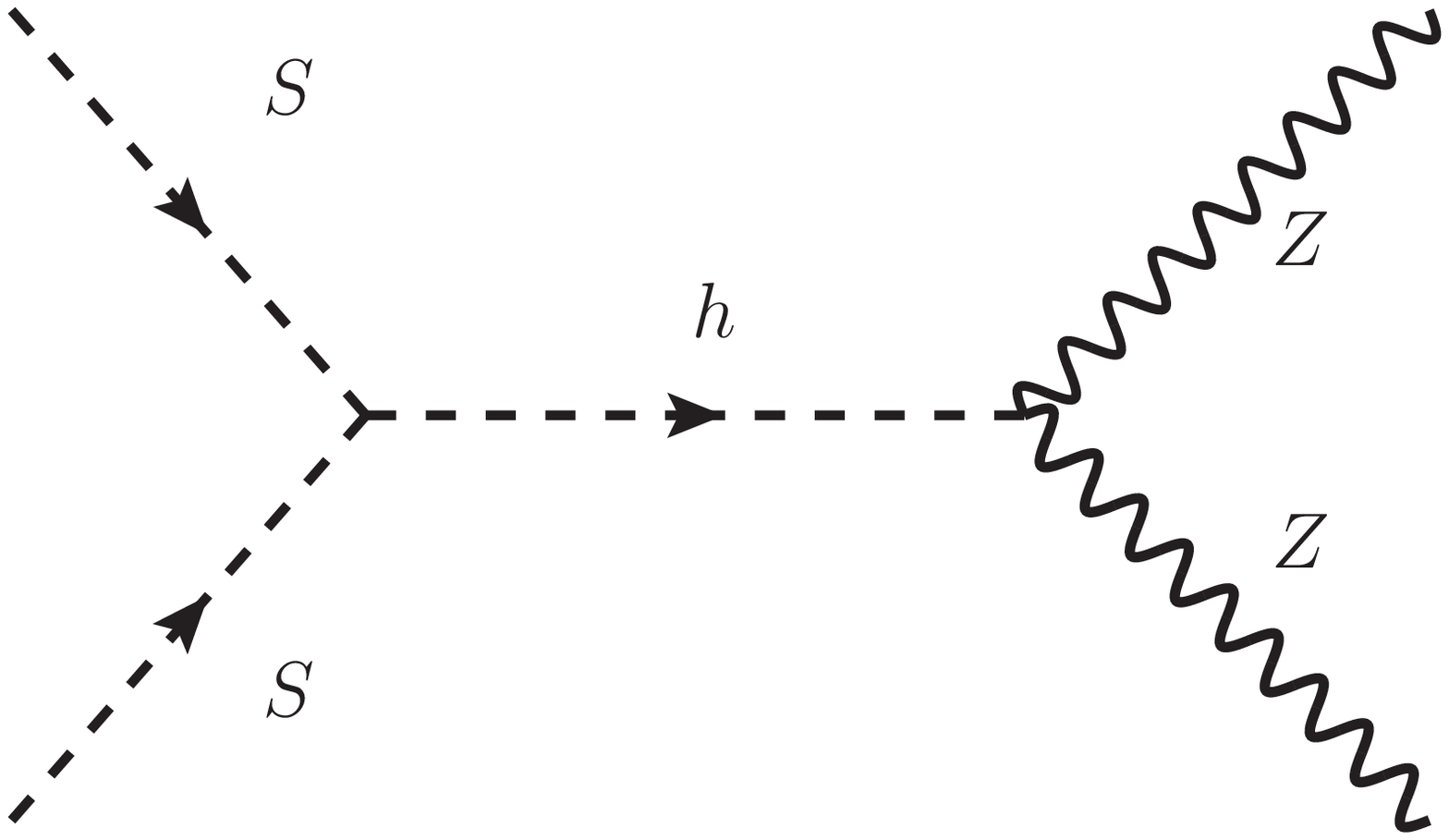}
\includegraphics[width=5cm,height=3cm]{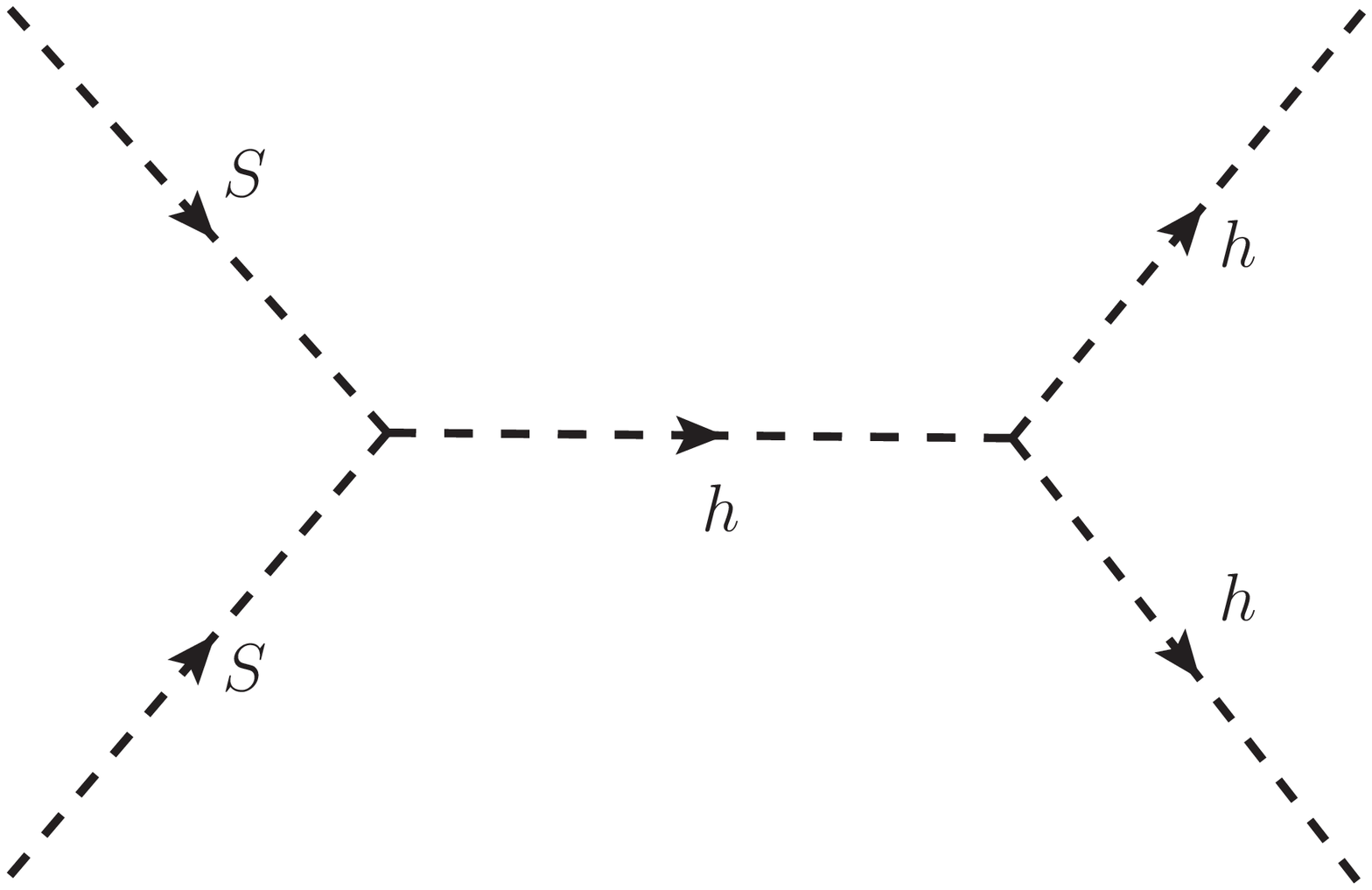}
\includegraphics[width=5cm,height=3.3cm]{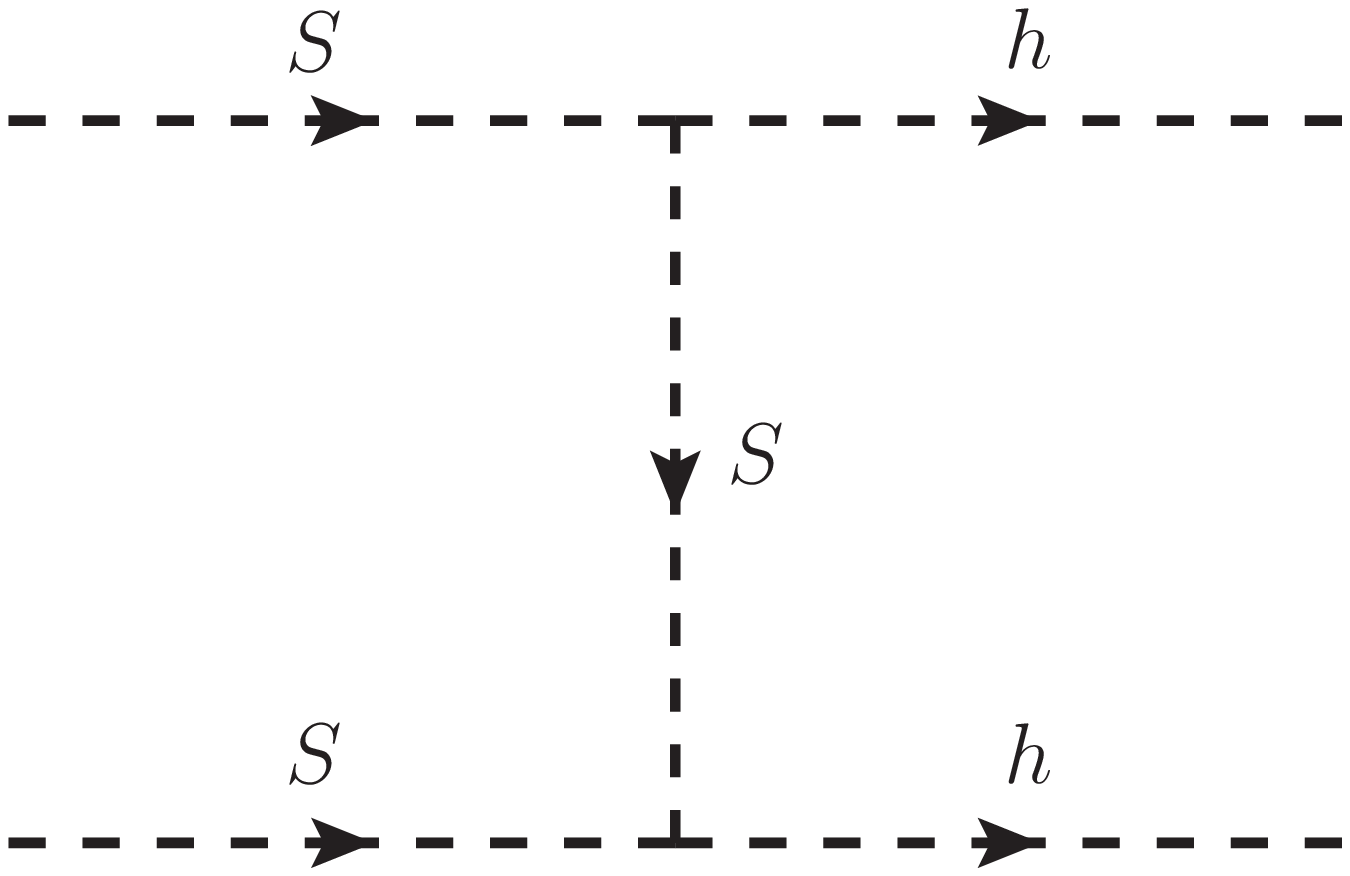}
\includegraphics[width=5cm,height=3cm]{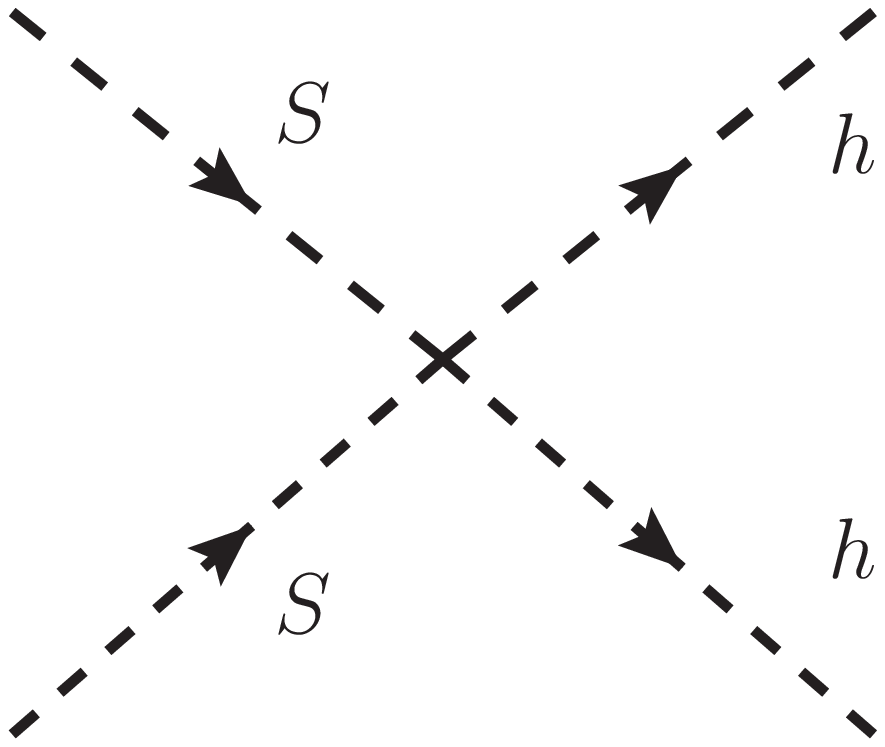}
\caption{Lowest order Feynman diagrams of two $S$ annihilate into a pair
of fermion and anti-fermion , $ W^+ W^-$, ZZ and Higgs.}
\label{anniS}
\end{figure}

In order to calculate the relic abundance of the scalar particle $S$ we 
have solved numerically the Boltzmann's equation which is given by
\bea
\frac{dn}{dt} + 3{\rm H}n = -\langle\sigma v\rangle (n^2-n_{\rm eq}^2)\,\, ,
\label{5}
\eea
where $n$ is the number density of the scalar particle $S$ and 
$n_{\rm eq}$ is the value of $n$ when $S$ was in  equilibrium 
(when temperature $T > T_f$,$T_f$ being the freeze out temperature of $S$), 
${\rm H}$ denotes the Hubble parameter, 
$\langle\sigma v\rangle$ is the thermal average of the product of
annihilation cross section and the relative velocity of the two
annihilating particles (in this case the scalar singlet $S$). 
It is useful to define two dimensionless quantities, 
$Y = n/s$ \cite{gondolo} and $x = m/T$. Where $s$ is the total entropy 
density of the universe, $T$ being the photon temperature. 
From the standard Friedmann-Robertson-Walker cosmology, 
the Hubble parameter ${\rm H} = \sqrt {\frac{8}{3}\pi G\rho}$ and 
$G$ is the gravitational constant. The total energy density ($\rho$) 
and the total entropy density ($s$) of the universe are given by
\cite{gondolo}
\bea
\rho &=& g_{eff}(T)\frac{\pi^2}{30}T^4 
\label{7a}
\eea
\bea
{\rm and}\,\,\,\,\, s &=& h_{eff}(T)\frac{2\pi^2}{45}T^3\,\,\, .  
\label{7b}
\eea
In Eqs.(\ref{7a}) and (\ref{7b}) $g_{eff}$, $h_{eff}$ are the effective 
degrees of freedom for the energy and entropy densities. 
Substituting Eqs. (\ref{7a}), (\ref{7b}) and the expression of ${\rm H}$ 
into Eq. (\ref{5}), we arrive at the equation for the evolution 
of $Y$ as
\begin{equation}
\frac{dY}{dx} = -\left( \frac{45}{\pi}G \right )^{-1/2}\frac{g_*^{1/2}m}{x^2}
\langle \sigma v \rangle (Y^2-Y_{eq}^2)\,\,\, ,
\label{8}
\end{equation}
where $g_*^{1/2}$ is defined as \cite{gondolo}
\begin{equation}
g_*^{1/2} = \frac{h_{eff}}{g_{eff}^{1/2}}
\left (1+\frac{1}{3} \frac{T}{h_{eff}}\frac{dh_{eff}}{dT} \right)\, .
\end{equation}
$Y_{\rm eq}$ is the value of $Y$ when $n = n_{\rm eq}$.
The expression for $Y_{eq}$ is given by \cite{gondolo}
\begin{equation}
Y_{eq} = \frac{45g}{4\pi^4}\frac{x^2K_2(x)}{h_{eff}(m/x)}
\label{9}
\end{equation}
where $g$ is the number of internal degrees of freedom of the
species under consideration (here $g$ = 1), $m$ is the mass 
and $K_n(x)$ is the modified bessel function of order $n$. 
From Eqs. (\ref{8}) and (\ref{9}), we obtain
\bea
\left( \frac{45}{\pi}G \right)^{-1/2}\frac{45g}{4\pi^4}
\frac{K_2(x)}{h_{eff}(T)}g_*^{1/2}m\langle \sigma v \rangle
\delta(\delta+2) 
= \frac{K_1(x)}{K_2(x)} -
\frac{1}{x}\frac{d \ln h_c(T)}{d \ln T}\; .
\label{10}
\eea
In the above equation, $h_c(T)$ is the contribution to $h_{eff}(T)$ from all 
species which are coupled to the universe at temperature $T$. 
Eq.(\ref{10}) above is solved 
numerically in a self consistent manner in order to obtain the value of $x_f$ 
(and hence the freeze out temperature $T_f$ ($= m/x_f$)). 
In the present case we have taken the value of $\delta$  
to be 1.5 \cite{gondolo}. 
Integrating 
Eq.(\ref{8}) from $x = x_0 = m/T_0$ to $x = x_f = m/T_f$, where $T_0$ 
is the present photon temperature which is of the order of $10^{-14}$ GeV 
($\sim0$), we obtain $Y_0$ (value of $Y$ at $T = T_0$). 
Knowing $Y_0$ we can compute the relic density of the
dark matter candidate (here $S$) 
from the relation \cite{gondolo},
\bea
\Omega h^2 &=& 2.755\times10^8 \frac{m}{\rm GeV}Y_0\,\,.
\label{11}
\eea
In the above $\Omega = \rho/\rho_c$ ($\rho$ being the dark matter density 
and $\rho_c$ is the critical density of the universe) and
$h = \frac {H} {100\,{\rm Km}\,\,{\rm sec}^{-1}{\rm Mpc}^{-1}}$.
Feynman diagrams in Fig. \ref{anniS} represent the possible annihilation channels of $S$. The expressions for annihilation cross section $\langle\sigma v\rangle$ for the processes like $SS\rightarrow f{\bar f}, W^+W^-, ZZ, hh$ are given in refs.\cite{mcdonald,lei}
\begin{figure}[h]
\centering
\includegraphics[width=7cm,height=8.5cm,angle=-90]{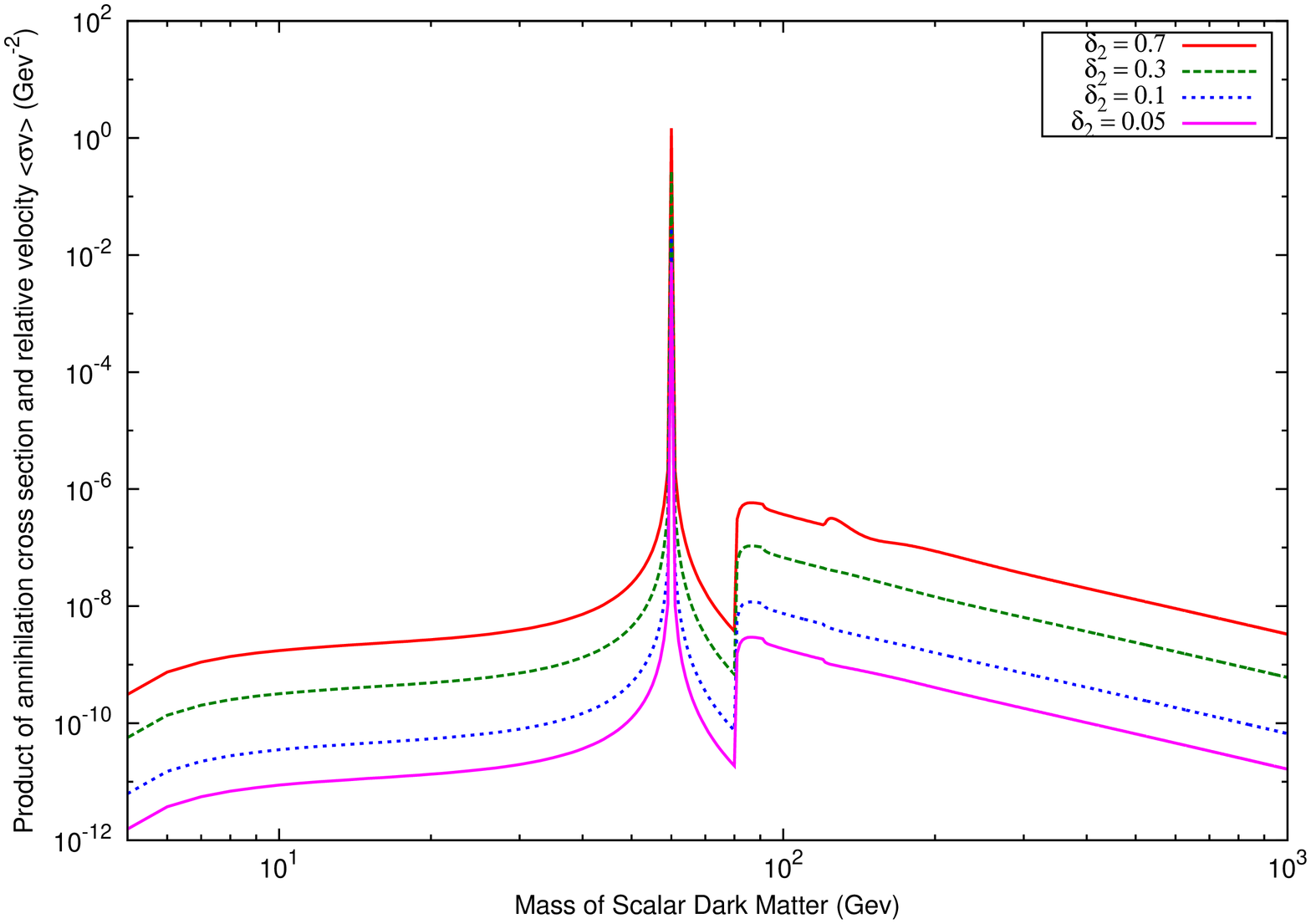}
\includegraphics[width=7cm,height=8.5cm,angle=-90]{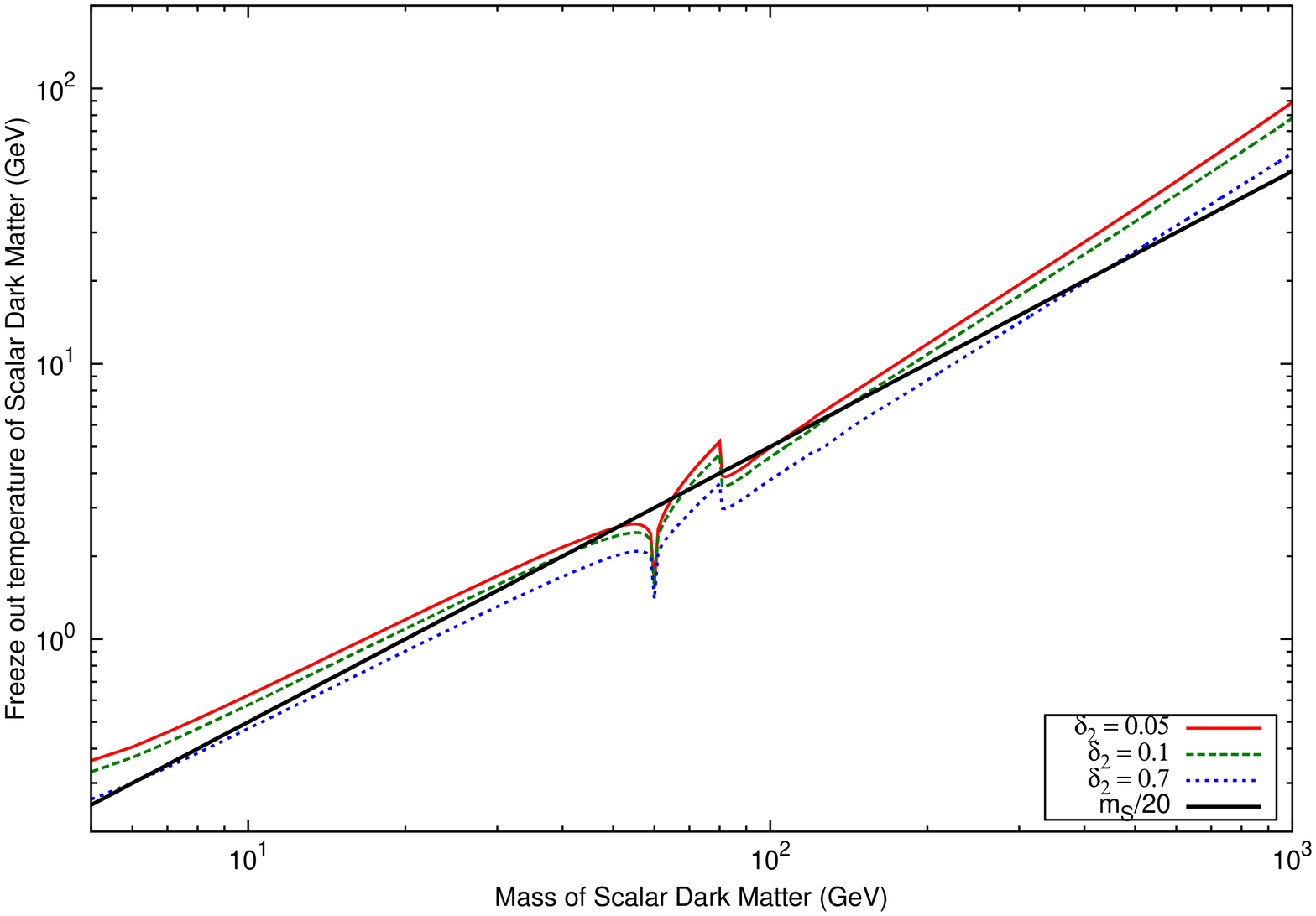}
\caption{Left Panel : Variation of product of annihilation cross section and 
relative velocity $\langle\sigma v\rangle$ with the mass of 
scalar dark matter $S$ for  $\delta_2$ = 0.7, 0.3, 0.1, 0.05, 
Right Panel : Variation of freeze out temperature $T_f$  
with the mass for different values of $\delta_2= 0.7,0.1,0.05$}
\label{cross}
\end{figure}
In this work we consider Higgs mass value $m_h = 120$ GeV. 
The variations of annihilation cross sections with scalar mass $m_S$ 
are shown in Fig. \ref{cross} (Left Panel) for different values of 
$\delta_2$. 
\section{Calculational Procedure and Results}
\begin{figure}[h]
\centering
\includegraphics[width=8cm,height=14cm,angle=-90]{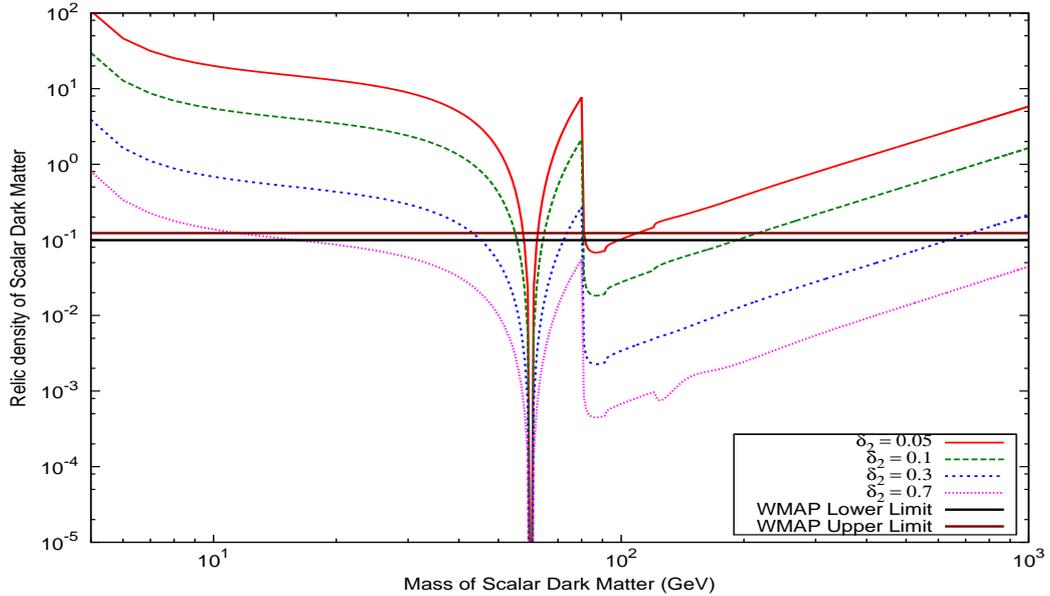}
\caption{Variation of relic density  $\Omega h^2$ with the mass of 
scalar dark matter for Higgs mass 120 GeV}
\label{density} 
\end{figure}

The relic density for scalar dark matter is obtained after an elaborate 
computation. We first calculate the freeze out temperature $T_f$ for 
scalar dark matter with different values of coupling constant 
$\delta_2$ and mass $m_S$. For this purpose we have solved Eq. (\ref{10}) 
numerically. The values of the quantities $g_*^{1/2}$, $g_{eff}^{1/2}$ 
and $h_{eff}$ for different $T$ required for solving Eq.(\ref{10}), 
are obtained from the figures (for the QCD phase transition 
temperature of 150 MeV) given in Refs. \cite{gondolo,olive}. 
In the Fig \ref{cross} (Right Panel), representative plots are 
the variations of $T_f$  in the scalar dark matter mass range 
5 GeV to 1 TeV for  different values of $\delta_2$ 
($\delta_2 $ = 0.05, 0.1, 0.7) and $m_h$ = 120 GeV 
with the topmost plot is for the lowest 
value of $\delta_2$ considered and the plots below are for 
other considered values of $\delta_2$ in the increasing order.  
In general, the freeze out temperature $T_f$ is approximately given by  
$T_f \sim m_S/20$. The plots for $T_f = m_S/20$ are also shown in  
Fig. \ref{cross} (Right Panel) (black dashed lines) for reference. 
The sudden dip in the values for $T_f$ in Fig. \ref{cross} (Right Panel)
around $m_S$ = 60 GeV can be understood from the expression of 
$\langle \sigma v \rangle_{f \bar f}$ (given in refs. \cite{mcdonald,lei}). 
At this point there is a 
sudden rise in $\langle \sigma v \rangle$ (Fig. \ref{cross} (Left Panel)) 
and it is due to Higgs propagator appearing in the annihilation process 
($SS\rightarrow f\bar f$). 
Using the values of freeze out temperatures (calculated from Eq. (\ref{10})) 
in Eq. (\ref{11}) the relic densities of the scalar dark matter 
for different scalar dark matter masses and different values of 
$\delta_2$ are computed. 
The results are shown in Fig. \ref{density}.  
In Fig. \ref{density} the two parallel lines denote the WMAP limits 
on relic density of dark matter ($0.099\leq \Omega h^2 \leq 0.123$). 
The different plots in Fig. \ref{density} 
correspond to different values of $\delta_2$ namely 
$\delta_2$ = 0.05, 0.1, 0.3, 0.7 
respectively with the topmost one is for smallest value of 
$\delta_2$ considered and 
the successive lower plots are for the other 
considered values of $\delta_2$ in increasing order. 
We have seen from Fig. \ref{cross} (Left Panel) that 
initially the annihilation cross section of $S$ increases 
with $m_S$; then at $m_S \approx m_h/2$, $\langle \sigma v \rangle$ 
rises rapidly and after which 
it decreases with the increase of $m_S$. Again for $m_S \sim 81$ GeV, 
$\langle \sigma v \rangle$ suddenly increases upto nearly 2 orders of 
magnitudes from its value at $m_S \sim $ 80 GeV and it is due to the fact that 
for $m_S >$ 80.4 GeV the annihilation channel $SS \rightarrow W^+W^-$
becomes kinematically possible. Thereafter $\langle \sigma v \rangle$ 
starts decreasing with the increase of $m_S$. Since relic 
density is inversely proportional to $\langle \sigma v \rangle$ \footnote{
Physically we can say that $\langle \sigma v \rangle$ is directly 
proportional to probability of that process. So for higher 
$\langle \sigma v \rangle$ the probability of pair annihilation 
of $S$ is high and hence density is low}, the variation 
of relic density of dark matter particle $S$ with $m_S$ is just 
opposite to the variation $\langle \sigma v \rangle$ with $m_S$. 
This feature is reflected Fig. \ref{density}.
Also since $\langle \sigma v \rangle$ is directly proportional 
to $\delta_2^2$ and its higher powers, 
higher the value of $\delta_2$ lower is the value of relic density 
(Fig. \ref{density}). 
\section{Constraining the model parameter $\delta_2$}
\begin{figure}[h]
\centering
\includegraphics[width=8cm,height=14cm,angle=-90]{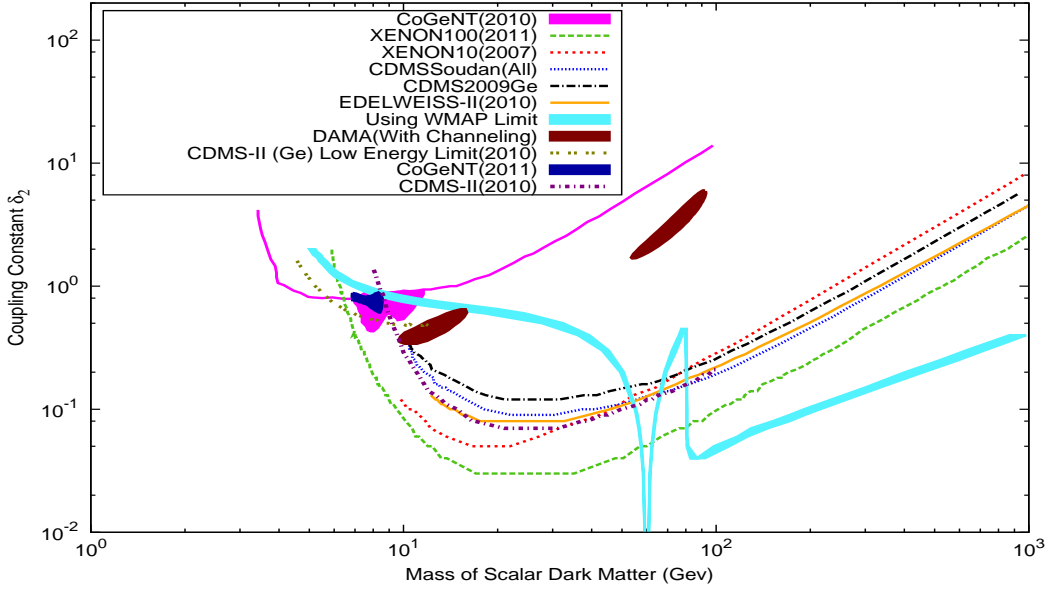}
\caption{Constraining the parameter space ($m_S - \delta_2$) of 
scalar dark matter for Higgs mass 120 GeV (upper panel) 
by using WMAP limit and recent experimental results of CDMS-II, 
XENON-10, XENON-100, CoGeNT, DAMA, EDELWEISS-II.}
\label{constrain}
\end{figure}

The model parameter $\delta_2$ is a very important parameter 
for this present model because it appears in both the expressions of
annihilation and scattering cross section of scalar dark matter $S$.
The spin independent scattering cross section for scalar dark matter
$S$ is given later in Eq. (\ref{14}).
In this section we have constrained the parameter space ($m_S
- \delta_2$) by using WMAP limit on relic density of dark matter and 
the results of recent experiments like CoGeNT, DAMA, CDMS-II, 
XENON-10, XENON-100 and  EDELWEISS-II. Similar to the previous 
discussions, here also we perform the calculations for Higgs mass 
$m_h$ = 120 GeV with 
$m_S$ in the range $5$ GeV $\leq m_S \leq 1$ TeV. The results  
obtained are shown in Fig. \ref{constrain}.
We first use the WMAP limit ($0.099\leq \Omega h^2 \leq 0.123$) 
on relic density of dark matter and using that limit we get 
the allowed values of $m_S$ for each value of $\delta_2$ 
(from Fig. \ref{density}). These results are shown in Fig. \ref{constrain} 
using turquoise coloured contour. 
Thereafter we estimate the allowed values 
of $\delta_2$ and $m_S$ using the mass - cross section limits given by the  
experiments like CDMS-II, DAMA, CoGeNT, XENON-10, XENON-100,
EDELWEISS-II and Eq. (\ref{14}).
In Fig. \ref{constrain} the magenta coloured contour 
represent the allowed regions of scalar dark matter $S$ 
obtained from CoGeNT data (2010). The overlap regions between these contours
(magenta and Turquoise) are therefore satisfied by both WMAP and CoGeNT 
(2010) results. From the overlap region (Fig. \ref{constrain}) 
the range of $m_S$ (in GeV) is found to be $7.7 \leq m_S \leq 11.15$ 
and the corresponding range of coupling $\delta_2$ is obtained as 
$0.7\leq \delta_2 \leq 0.95$. These ranges 
therefore satisfy both WMAP and CoGeNT (2010) limits. 
The dark blue coloured contours indicate (Fig. \ref{constrain}) 
new bounds from CoGeNT data (2011) \cite {cogent11}. 
The common region between CoGeNT (2011) and 
WMAP lies in the range $8.0 \leq m_S \leq 8.53$ GeV,
$0.8\leq \delta_2 \leq 0.9$. 
This common intersection region is also well supported by CoGeNT (2010)
and CDMS-II (2010) data \cite {simple} (purple dashed line).  

Similar $m_{S} - \delta_2$ contours obtained from the 
DAMA experiment results (with channeling) 
are shown as maroon contours in Fig. \ref{constrain}. 
One sees that the small overlap regions between the two 
contours (turquoise and maroon) are restricted by the scalar mass 
(in GeV) range $14.8\leq m_S \leq 15.9$. 
The corresponding values of $\delta_2$ are found around 0.6.
We remark in the passing that we have checked for other allowed regions
in ``dark matter mass $ - \sigma^{\rm scalar}_{\rm nucleon}$" 
plane given by DAMA experiment 
but we have not obtained any overlap region such as described above. One 
of such regions (maroon coloured contour) is also shown in Fig. 
\ref{constrain}. The olive dashed line in Fig. \ref{constrain} 
represent upper bounds that we have obtained from low energy analysis 
of the CDMS-II Germanium data \cite {cdms2-10}. But it has no intersection 
with WMAP satisfied region (turquoise coloured contour).

But unlike CoGeNT and DAMA, other experiments like XENON-10, XENON-100, 
CDMS-II, EDELWEISS-II do not provide a bounded allowed region 
in $m_S$ - $\sigma_{\rm N}$ plane
($\sigma_{\rm N}$ is the scattering cross section of dark matter 
and nucleon). Instead they provide  upper bounds of scattering cross section 
for a particular mass of dark matter. Consequently we also obtain 
upper bounds of $\delta_2$ for a specific mass of $S$ for those experiments. 
These results
are also shown in Fig. \ref{constrain}. In Fig. \ref{constrain} 
green dashed line represents the upper bound of $\delta_2$ for 
XENON-100. CDMS-II results (CDMS 2009 Ge and CDMS Soudan (All) \footnote 
{Which are the results obtained by the CDMS-II collaboration from the 
combined analysis of full data set of Soudan.}) are shown by black and 
blue dashed lines. XENON-10 and EDELWEISS-II results are represented 
by red dashed, orange solid line respectively.
The WMAP results (turquoise plot in Fig. \ref{constrain})
intersect with the upper bounds obtained from 
XENON-10 results in $m_S - \delta_2$ plane are found to be at 
the values of $m_S = 53.5$, $\delta_2 = 0.12$, 
$m_S = 67.4$, $\delta_2 = 0.16$ and $m_S = 80.2$, 
$\delta_2 = 0.21$. Therefore the turquoise region below 
($m_S = 53.5$, $\delta_2 = 0.12$), ($m_S = 67.4$, $\delta_2 = 0.16$) 
and ($m_S = 80.2$, $\delta_2 = 0.21$) is obeyed by both WMAP and XENON-10. 
Similarly the regions satisfied by both CDMS-II results
(CDMS 2009 Ge, CDMS Soudan (All)) and 
WMAP are represented by the turquoise colour below intersection points      
($m_S = 52.0, \delta_2 $= 0.15), ($m_S = 67.9$, $\delta_2 = 0.17$),
($m_S = 80.3$, $\delta_2 = 0.20$) and ($m_S = 53.9, \delta_2 $= 0.12),
($m_S = 66.5, \delta_2 $= 0.14), ($m_S = 80.4, \delta_2 $= 0.16) 
respectively as shown in the Fig. \ref{constrain}.
Also the overlap regions of WMAP, XENON-100 and WMAP, EDELWEISS-II 
are below the following intersection points, 
can be read out from the Fig. \ref{constrain} 
as ($m_S = 57.6$, $\delta_2 = 0.05$), ($m_S = 62.5$, $\delta_2 = 0.05$), 
($m_S = 80.5$, $\delta_2 = 0.07$) and ($m_S = 54.1$, $\delta_2 = 0.11$), 
($m_S = 66.7$, $\delta_2 = 0.14$), ($m_S = 80.4$, $\delta_2 = 0.17$).
We have also found that for XENON-100 there is another intersection point 
with WMAP in the lower mass region around $m_S \sim$ 6.0 GeV, $\delta_2
\sim$ 1.25.  
 
From the above analyses it appears that there are two distinct regions 
in the $m_S - \delta_2$ plane for scalar dark matter $S$ which are 
allowed regions for both WMAP and recent experiments. The regions can 
be classified as follows.
\begin{itemize}
\item A lower mass region where we have found 3 mass ranges for
scalar dark matter $S$. These ranges are given by $m_S \sim$ 6 GeV
($\delta_2 \sim$ 1.25), 7.7 GeV $\leq m_S \leq$ 11.15 GeV
($0.7 \leq \delta_2 \leq 0.95$) 
and 14.8 GeV $\leq m_S \leq$ 15.9 GeV ($\delta_2 \sim$ 0.6).
The corresponding ranges for 
coupling $\delta_2$ which we have found are given within brackets.
This lower mass domain is supported by WMAP and
various ongoing dark matter direct detection experiments. In
this case $m_S \sim $ 6 GeV is supported by WMAP and XENON-100. Second and 
third mass ranges are obeyed by WMAP, CoGeNT (2010) data and WMAP, 
DAMA (with channeling) data respectively (Fig. \ref{constrain}). 

But if we use more recent data of CoGeNT (CoGeNT (2011) data) then
the second mass range of scalar dark matter $S$ gets reduced to 
8.0 GeV $\leq m_S \leq$ 8.53 GeV. The ranges for coupling 
$\delta_2$ also reduced to $0.8 \leq \delta_2 \leq 0.9$. 
It is also seen from Fig. \ref{constrain}
that is region is supported by CDMS-II (2010) bounds. Other mass
ranges are remain unchanged.
\item A higher mass region with the scalar dark matter mass
range $\,\sim 52.5 {\rm GeV} \leq m_S \leq \,\sim 1000$ GeV,
with the range of $\delta_2$ found as
$0.02 \leq \delta_2 \leq 0.4$ (Fig. \ref{constrain}).  
This mass region is satisfied by the allowed domains of 
WMAP, CDMS-II, EDELWEISS-II, XENON-10 and XENON-100.
\end{itemize}
  
Here we make some comments about the region of the parameter
space ($m_S$ vs $\delta_2$) that is not satisfied by the results of
direct detection experiments we have considered. In this region (15 GeV$<m_S<$ 52.5 GeV) the values of $\delta_2$ required to obtain current relic density
(within WMAP limit) are such that the scattering cross sections obtained using these values for different dark matter masses do not satisfy the experimental limits given by different exclusion plots. 
\section{Formalism for the Calculations of Direct Detection Rates} 
\begin{figure}[h]
\centering
\includegraphics[width = 6cm,height = 4cm]{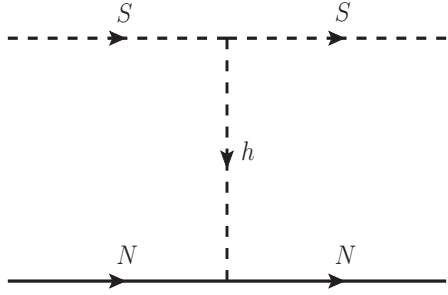}
\caption{Feynman diagram for the elastic scattering between $S$ 
and nucleon $N$ via Higgs exchange.}
\label{feynman}
\end{figure}

In this section we estimate the differential direct detection rates 
and their annual variations for scalar dark matter 
$S$. For this purpose we have chosen $^{76}$Ge and $^{131}$Xe 
as a detector materials.
The direct detection of dark matter by a terrestrial detector 
uses the principle of elastic scattering of dark matter particles 
off the detector nuclei and the energy of the recoil nucleus is measured.  
It is very difficult to measure the low recoil energy of nuclei 
accurately and hence a very low threshold and 
low background detector is required. In Fig. \ref{feynman} we show the Feynman 
diagram for such elastic scattering process of scalar singlet $S$ 
through Higgs exchange. The 
scalar singlet $S$ - nucleon $N$ elastic scattering ($SN\rightarrow SN$) 
cross section \cite{Burgress} is given by
\begin{eqnarray}
\sigma^{\rm scalar}_{N} 
&=&
\frac{\delta_2^2 v^2 |{\cal A}_N|^2}{4\pi} 
\left( \frac{m^2_r}{{M_S}^2{M_h}^4}\right)\,\, ,
\label{14}
\end{eqnarray}
where, $m_r (N,S)= M_N M_S/(M_N + M_S)$ is the reduced mass, $\cal A$ 
is coupling between Higgs and nucleon $N$ and its value is $\sim$
340 MeV/{$V$} \cite{Burgress}, with $V$ being the VEV of 
Higgs boson. The scalar 
singlet - nucleus elastic scattering cross section is given by \cite {Burgress} 
\begin{eqnarray}
\sigma^{\rm scalar}_{\rm nucleus} 
&=& \frac{A^2 m^2_r({\rm nucleus},S)}{ m^2_r({\rm nucleon},S)} 
\sigma^{\rm scalar}_{\rm nucleon}\,\, .
\label{15}
\end{eqnarray}
In the above $A$ is the mass number of the nucleus.
The differential detection rate of dark matter per unit detector 
mass is given by \cite{jungman} 
\bea
\frac {dR} {dE_R} 
&=& \frac {\sigma^{\rm scalar}_{\rm nucleus}\rho_S} {4v_e m_S
m_r^2} F^2 (E_R) \nonumber \\
&& \times \left[
{\rm erf}\left(\frac{v_{min} + v_e}{v_0}\right) 
- {\rm erf}\left(\frac{v_{min} - v_e}{v_0}\right)
\right]\,\, 
\label{23} \\
=&&T_1 T_2 T_3 \nonumber\,\,\,\,\,\,\,\,\,\, ,
\eea
where 
\bea
&&T_1 = \frac {\sigma^{\rm scalar}_{\rm nucleus}\rho_S} {4v_e m_S
m_r^2},\,\, 
T_2 = F^2 (E_R),\,\, \nonumber \\ 
&&T_3 = \left[
{\rm erf}\left(\frac{v_{min} + v_e}{v_0}\right)
- {\rm erf}\left(\frac{v_{min} - v_e}{v_0}\right)
\right]\,\, .
\label{23a}
\eea
Where ${v}_e$ is the velocity of earth with respect to galactic frame 
of reference. And its expression is given by \cite{jungman},
\bea
v_e &=& v_\odot + v_{\rm orb} \cos\gamma 
\cos \left (\frac {2\pi (t - t_0)}{T} \right ) \; . 
\label{20}
\eea
In the above expression $t$ denotes any time of the year, 
$T = 1$ year is the time period of earth's motion
around the sun, $v_{\rm orb}$ = 30 Km/sec is earth's orbital speed and 
$\gamma \backsimeq 60^0$ is the angle subtended by the ecliptic 
at the galactic plane. The solar velocity  $v_\odot$ is given 
by  
\bea
v_\odot &=& v_0 + v_{\rm pec}\,\, ,
\eea
where $v_0$ is the circular speed of sun
around the galactic centre taken to be 220 Km/sec 
and $v_{\rm pec}$ is the peculiar velocity with $v_{\rm pec} = 12$ Km/sec.
The periodicity in Eq. (\ref{20}) causes an annual modulation 
of the event rates of dark matter in a terrestrial detector 
which serve as a definite signal of dark matter detection. In the 
Eq. {\ref{23}} $F(E_R)$ is the nuclear form factor given by \cite{engel},
$\rho_S$ is the dark matter density in 
the solar neighbourhood, equal to 0.3 GeV/cm$^3$ for the rest of 
our calculations in this section . 
$v_{min}$ denotes the minimum velocity of dark matter required 
to produce a recoil energy $E_R$. The expression of $v_{min}$ is
given by,
\bea
v_{min} = \left(\frac{m_{nucleus}E_R}{2m_r^2}\right)^{1/2} \;. 
\label{19}
\eea

The measured response of the detector by the scattering of dark matter
off detector nucleus is in fact a fraction of the actual recoil energy. 
Thus, the actual recoil energy $E_R$ is quenched by a factor 
$q_X$ (different for different
nucleus X) and we should express differential rate in Eq. (\ref{23}) 
in terms of $E = q_X E_R$.
Thus the differential detection rate (events/Kg/Day/keV) in terms of the 
observed recoil energy $E$ for a monoatomic detector like Xe 
can be expressed as
\begin{equation}
\frac {\Delta R} {\Delta E} (E) =
\int^{(E + \Delta E)/q_{\rm Xe}}_{E/q_{\rm Ge}}
\frac {dR_{\rm Xe}} {dE_R} (E_R) \frac {dE_R} {\Delta E} \,\, .
\label{24} 
\end{equation}
The total detection rate of dark matter is obtained by integrating 
Eq.(\ref{23}) as 
\bea
R = \int_{E_T}^{\infty} \frac {dR}{dE_R}dE_R \,\, ,
\label{26}
\eea
where $E_T$ is the threshold energy for a given detector material.
\section{Direct Detection Rates for Scalar Dark Matter}
In the present work, computations of direct detection rates are 
performed with $m_h = 120$ GeV, $\Delta E = 0.5$ keV and at a time 
$t = t_0$. 
As discussed earlier we have computed the direct detection 
rates and their annual variations for each of the detector materials 
namely $^{76}$Ge, $^{131}$Xe. The quenching factors for 
$^{76}$Ge = 0.25 \cite {bottino}, $^{131}$Xe = 0.8 \cite {bottino}, 
The differential detection rates and their annual variations can now 
be computed using Eqs. (\ref{23}) - (\ref{26}). 
\begin{figure}[h]
\centering
\includegraphics[width=6cm,height=8cm,angle=-90]{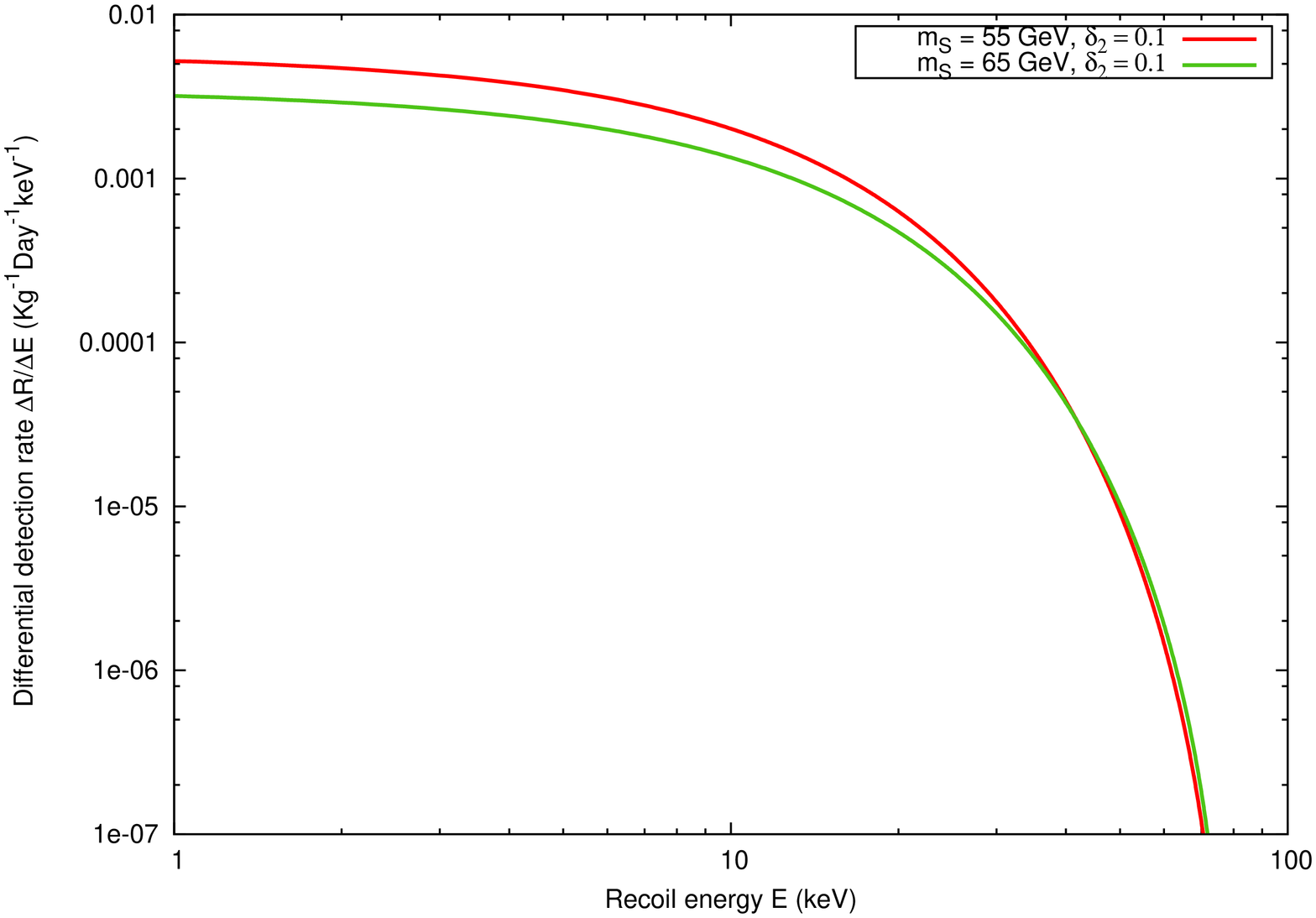}
\includegraphics[width=6cm,height=8cm,angle=-90]{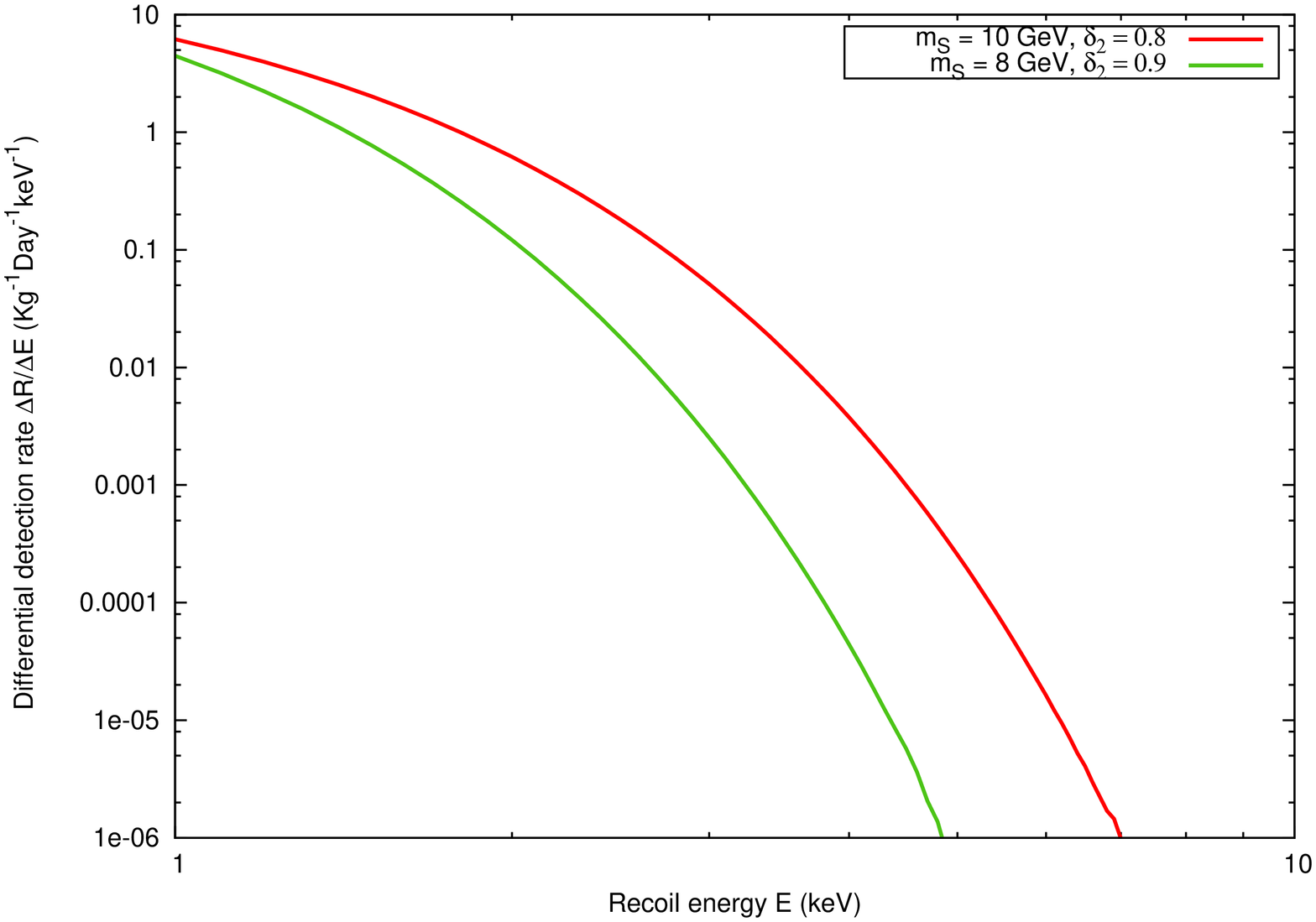}
\caption{Variation of differential detection rates $\Delta R/\Delta E$
of scalar dark matter $S$ with observed recoil energy E for monoatomic
detectors Xe (left panel), Ge (right panel)}
\label{mono}
\end{figure}

The variation of differential detection rates of scalar dark matter $S$ 
with observed recoil energy E for mono atomic 
targets like Xe, Ge are shown in Fig. \ref{mono}. 
In the left panel of Fig. \ref{mono}, we 
show the estimates of differential detection rates for different 
values of observed recoil energy $E$, with Xe as target material 
for $m_S = 55$ GeV (red solid line) and $m_S = 65$ GeV (green solid line). 
For both the cases the value of the coupling constant $\delta_2$ 
is taken to be 0.1 (in agreement with the higher mass region 
described in section 5). Left panel of Fig. \ref{mono} shows 
that although the two plots corresponding to two scalar masses 
are distinguishable at lower recoil energies 
($\leq 11$ GeV), at higher recoil energies they tend to coincide. 
In the right panel of Fig. \ref{mono} we show the direct detection 
rates results for the case of Ge. In this case, calculations are 
performed for two sets of $m_S - \delta_2$ values namely 
($m_S = 10$ GeV, $\delta_2 = 0.8$), ($m_S = 8$ GeV, $\delta_2 = 0.9$).
These values are chosen from the allowed lower mass domain discussed 
in section 5 (Fig. \ref{constrain}). It is seen from the right
panel of Fig. \ref{mono} that the rates for the set 
($m_S = 8$ GeV, $\delta_2 = 0.9$, represented by green solid line), 
falls off faster than those for the set 
($m_S = 10$ GeV, $\delta_2 = 0.8$, represented by red solid line).   
\begin{figure}[h]
\centering
\includegraphics[width=6cm,height=8cm,angle=-90]{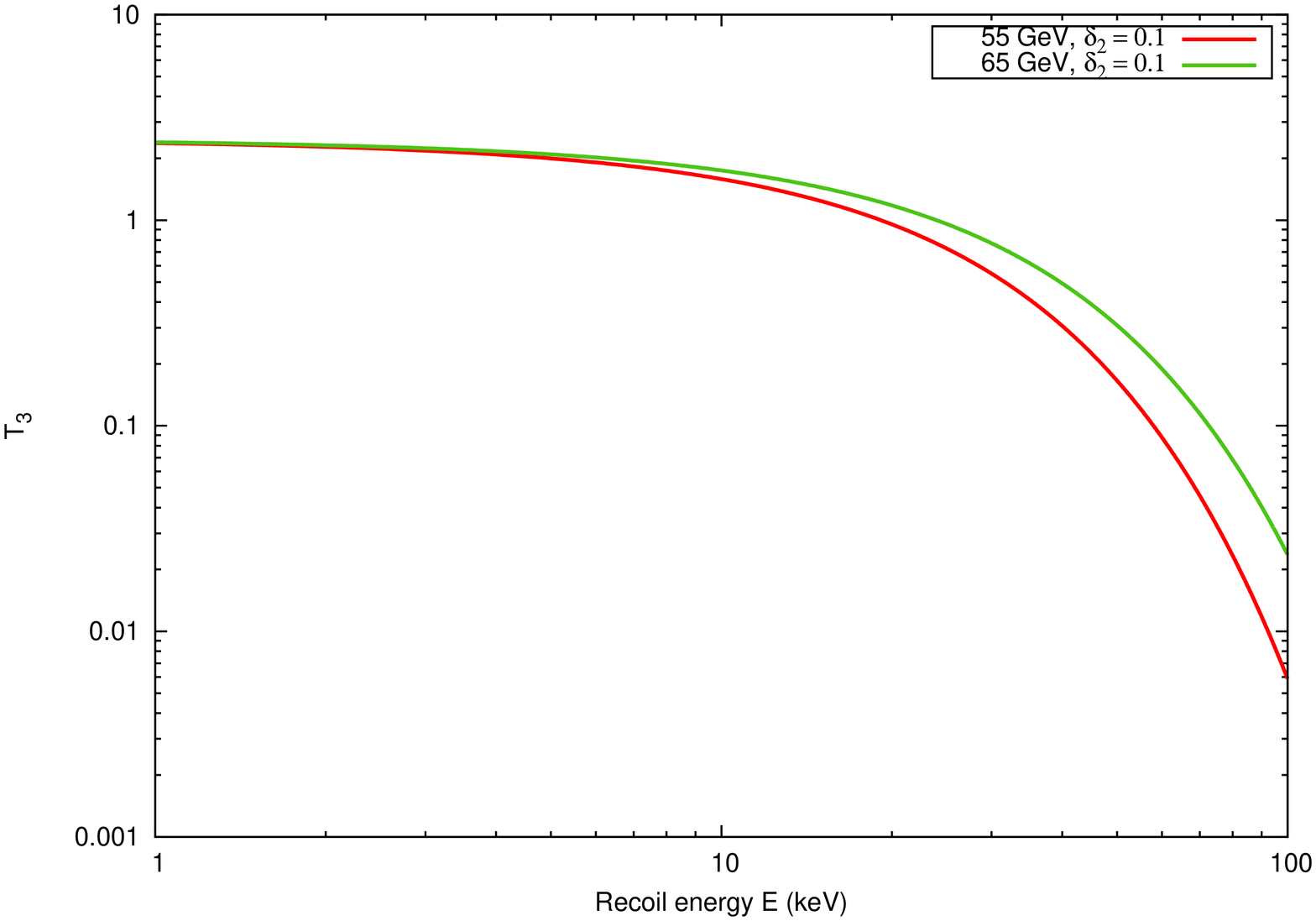}
\includegraphics[width=6cm,height=8cm,angle=-90]{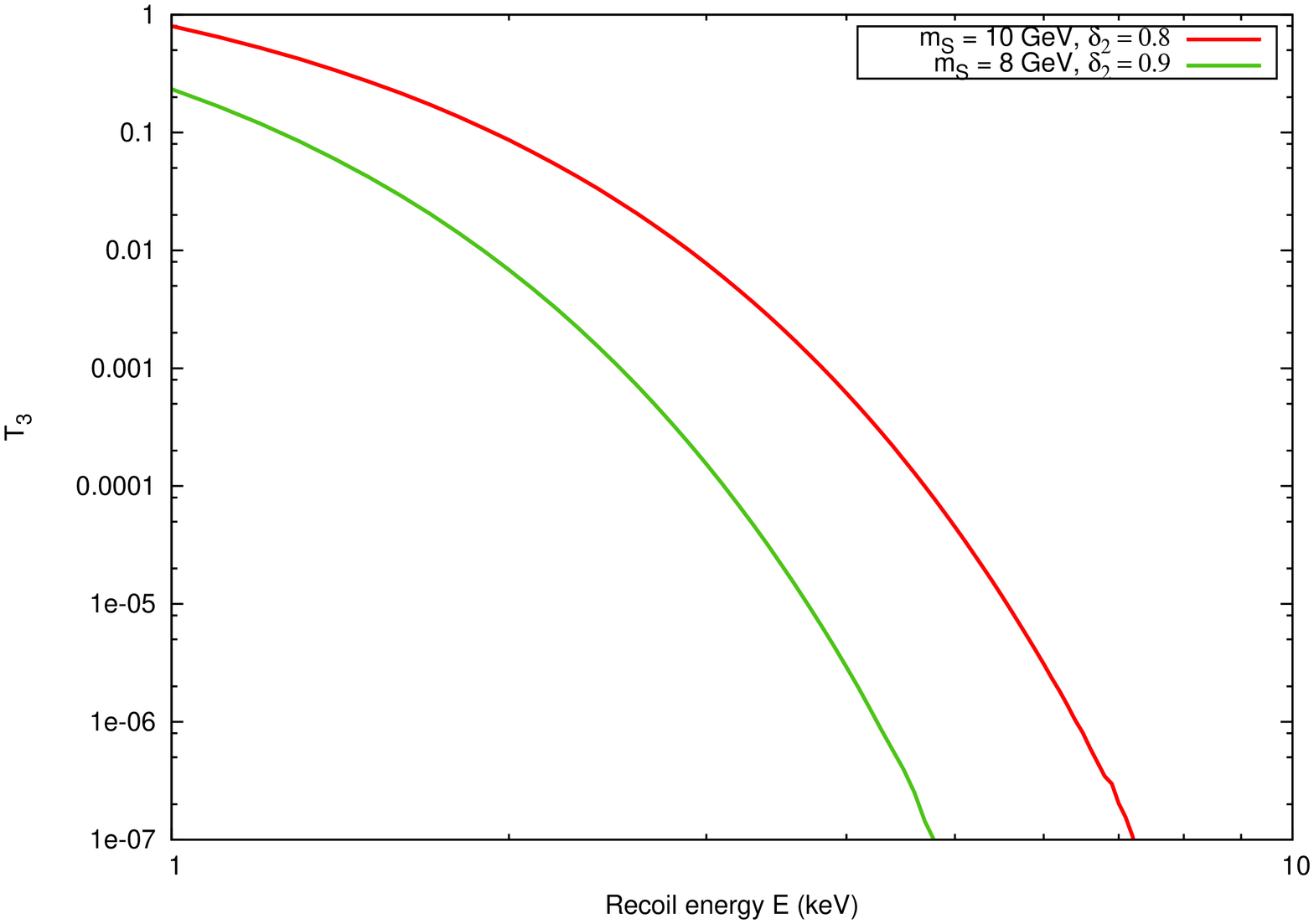}
\caption{Left panel - Variation of $T_3$ with observed recoil
energy $E$ for Xe
with $m_S$ = 55 GeV (green solid line), 65 GeV (red solid line).
Right panel - Same for Ge with $m_S$ = 10 GeV (red solid line),
8 GeV (green solid line)}
\label{t3}
\end{figure}
The nature of $\Delta R/\Delta E$ for the case of Xe 
(left panel of Fig. \ref{mono}) can be explained by examining 
the variation of $T_3$ (Eq. \ref{23a})with $E$. This is shown 
in  left panel of Fig. \ref{t3}. 
For low values of $E$ and for high mass range of $S$ ($m_S \gtrsim $ 
55 GeV), $v_{min}$ (Eq. \ref{19})$\ll$ $v_e$ and hence $T_3$ is effectively 
independent of $E$. Therefore as $E$ increases, the values of 
$T_3$ for $m_S$ = 65 GeV becomes larger than those for $m_S$ = 55 GeV. 
Also $\sigma^{\rm scalar}_{\rm nucleus}$ is inversely proportional 
to $m_S$ and $T_1$ is directly proportional to $\sigma^{\rm scalar
}_{\rm nucleus}$. Consequently $T_1$ is inversely proportional to $m_S$.
Now the variation of $\Delta R/\Delta E$ 
with $E$ is due to the combined effects of both $T_1$ and $T_3$ 
(Eq. (\ref{23})). This explains the nature of the plots for Xe 
in Fig. \ref{mono}.  
In the case of Ge however, $T_1$ is nearly the same for both the 
masses considered. Consequently the variations of $\Delta R/\Delta E$ 
with $E$ (Fig. \ref{mono} (right panel)) is dominated only by the nature 
of variations of $T_3$ with $E$. This variations are shown in 
right panel of Fig. \ref{t3} which explains the nature of variations 
of $\Delta R/\Delta E$ with $E$ for Ge.

The annual variations of total detection rates of WIMP 
is a crucial evidence for dark matter. This 
variation is caused by the periodic motion of earth around 
the sun in which the directionality of earth's motion changes over the year. 
Since the solar system moves towards the direction of Cygnus constellation,
earth experiences a WIMP wind apparently coming from the direction of Cygnus.
Due to the periodic motion of earth,  the relative speed between earth and WIMP 
changes over the year. It becomes maximum when both the velocities 
of solar system and earth are in the same direction (on $2^{nd}$ June) 
in which case the earth encounters maximum WIMP flux. The WIMP flux 
encountered by the earth is minimum when velocities of earth and 
sun are in opposite direction. Consequently, 
maximum events are expected on $2^{nd}$ June of every year.

\begin{figure}[h]
\includegraphics[width=6cm,height=8cm,angle=-90]{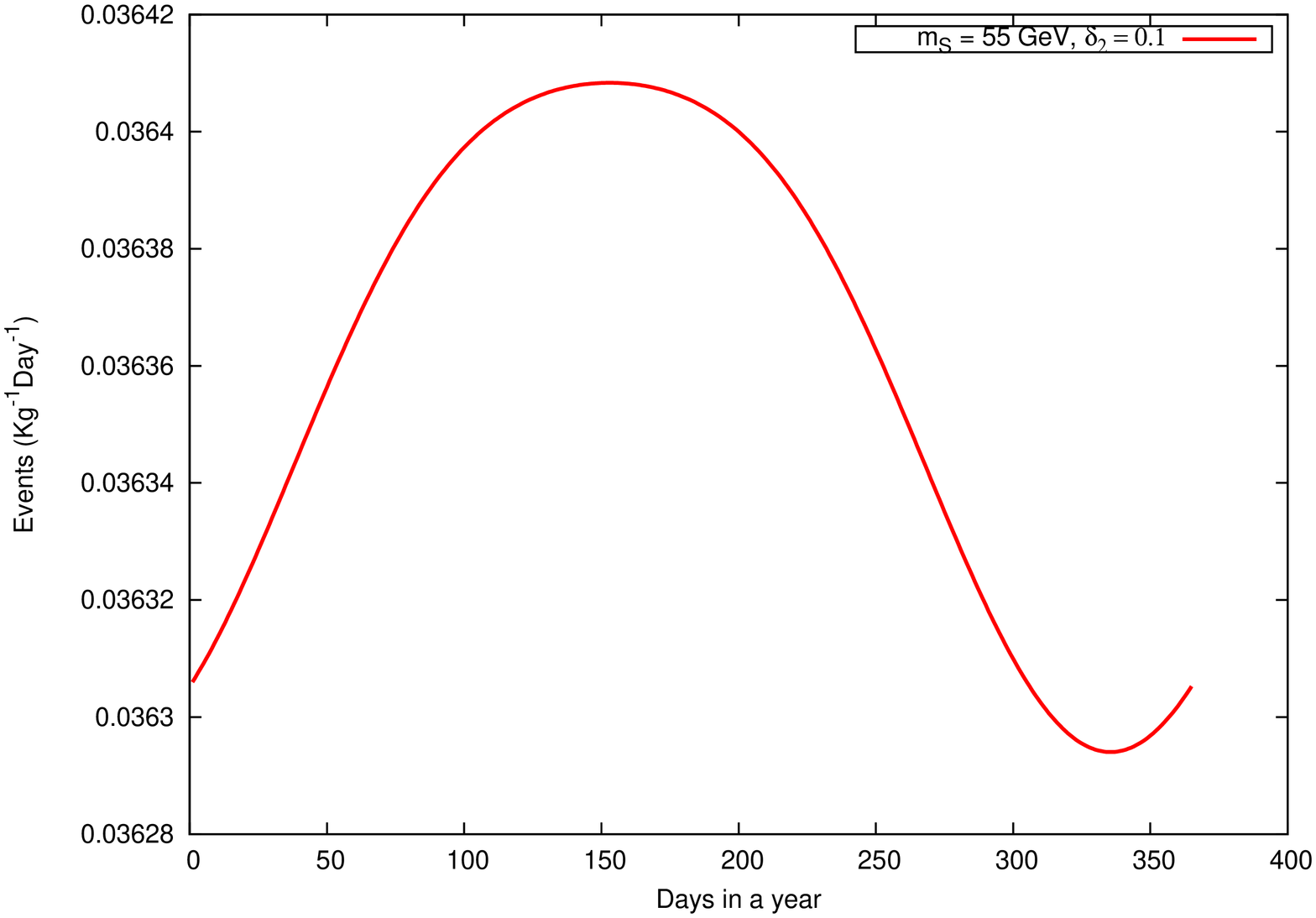}
\includegraphics[width=6cm,height=8cm,angle=-90]{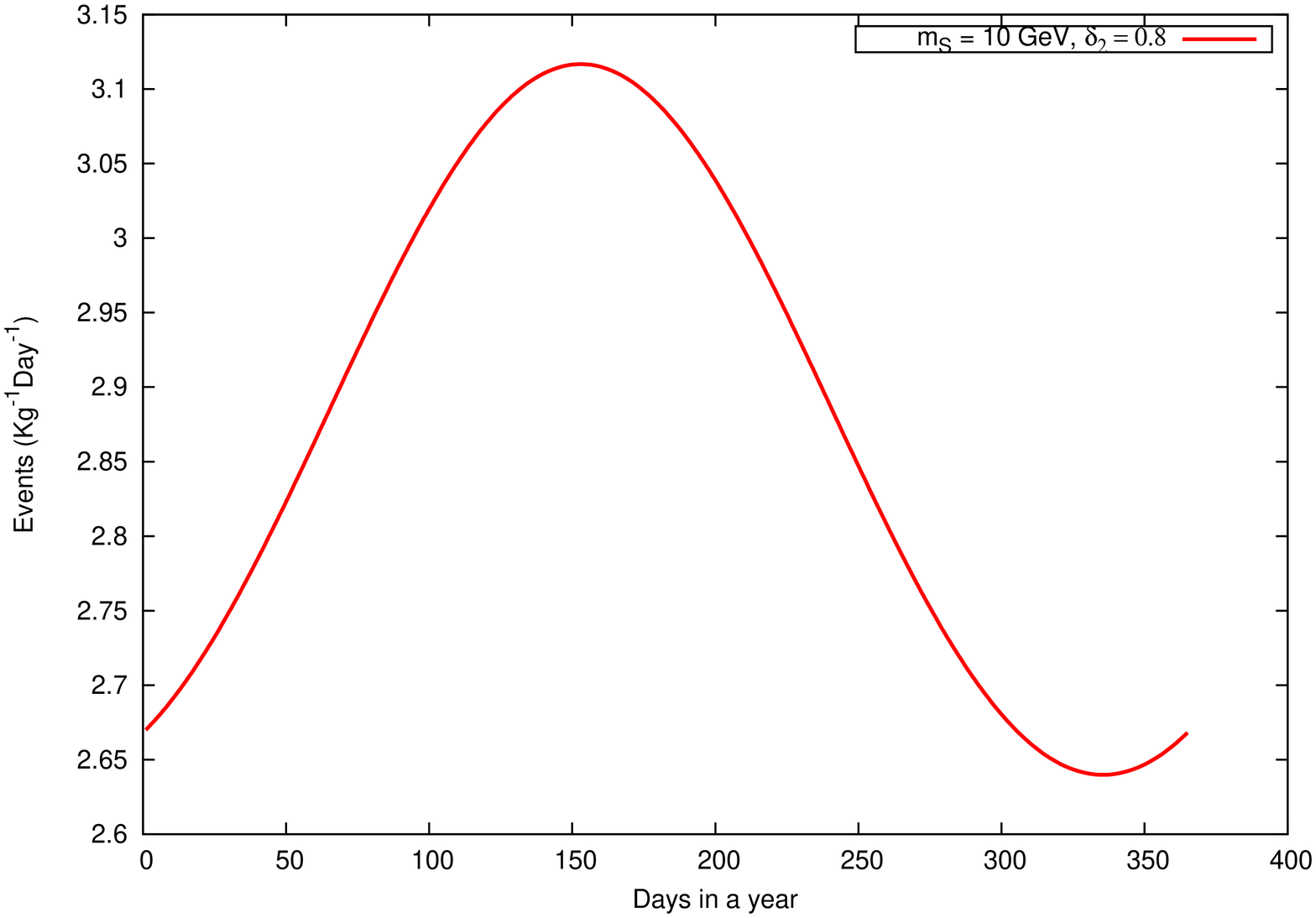}
\caption{Left panel - Annual variation of total detection rates of
scalar dark matter for Xe (mono atomic target) with $m_S$ = 55 GeV,
$\delta_2$ = 0.1. Right panel - Same for Ge with $m_S$ = 10 GeV,
$\delta_2$ = 0.8}
\label{annual}
\end{figure}

In this work we compute the total detection rates at each day
of a year, for the same detector materials namely Xe, Ge  
considering the scalar singlet as dark matter candidate. 
The results are then plotted with the days of year 
which show the annual variation of total detection rates.  
The calculations for Xe are performed for the set 
($m_S = 55$ GeV, $\delta_2 = 0.1$) whereas for Ge, 
the set ($m_S = 10$ GeV, $\delta_2 = 0.8$). 
The results for Xe and Ge are shown in left and right panels of 
Fig. \ref{annual} respectively. 
All the plots in Figs. \ref{annual} show that the 
maximum expected events are at $t = 153$ (day) (corresponds on $2^{nd}$ June).
\section{Indirect Detection of Scalar Dark Matter}
Another promising method for the detection of dark matter (WIMPs) is the
observation of annihilation products of dark matter present in the galactic
halo. In this section we will consider $\gamma$-rays coming from the
dark matter annihilation in the galactic centre (GC).

Recently it has been reported that there is a 4.6$\sigma$ (3.3$\sigma$)
\cite{gammaref1, gammaref2} local (global) evidence of a monochromatic 
gamma-ray line with an energy $E_{\gamma} \approx$ 130 GeV 
by the publicly available data \cite{fermipublic} of Fermi Large Area Telescope (Fermi-LAT).
This signal comes from two regions one of 
which is nearly at the centre of our galaxy ($-1^0, -0.7^0$),
hereinafter referred to as 
the ``central region" and another is located at ($-10^0, 0^0$), 
called the ``west region". Both regions are extended with in a circle of radius 
of $3^0$. It is suggested that this excess of gamma ray signal from galactic centre (GC) is not
associated with the Fermi bubbles \cite{gammaref2} and  may result from 
dark matter annihilation into two monochromatic gamma-rays.

\begin{figure}[h]
\centering
\includegraphics[width=7cm,height=3.5cm,angle=0]{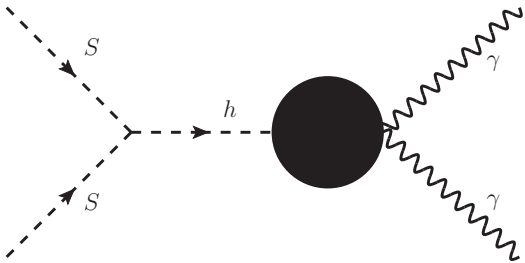}
\caption{Feynman diagram for the process $SS\rightarrow \gamma \gamma$}
\label{gamma}
\end{figure}

We have calculated the $\gamma$-rays flux due to 130 GeV scalar dark matter
annihilation in the ``central region" of our Milky way galaxy. The Feynman
diagram for the process $SS\rightarrow \gamma \gamma$ is shown in
Fig. \ref{gamma}.
The expression of $\gamma-$ray flux due to dark matter annihilation in 
galactic halo is given by \cite{gammafermi},
\bea
\frac{d{\Phi}_{\gamma}}{dE_{\gamma}} = \frac{1}{8\pi}
\frac{{\langle\sigma v \rangle}_{SS\rightarrow\gamma \gamma}}{m_S^2}
\frac{dN_{\gamma}}{dE_{\gamma}}r_{\odot}\rho_{\odot}^2J \,\, ,
\eea
where 
\bea
J = \int db \int dl \int_{l.o.s} \frac{ds}{r_\odot}\cos b
\left(\frac{\rho(r)}{\rho_\odot}\right)^2
\label {j} 
\eea
and
\bea
\frac{dN_{\gamma}}{dE_{\gamma}} = 2\delta(E-E_{\gamma})\,\, . 
\eea
In the above, $l$ and $b$ are the galactic longitude and latitude respectively.
We have performed $l$, $b$ integration (in Eq. \ref{j}) over the ``central
region" of the our galaxy and the $s$ integration (in Eq. \ref{j}) along the 
line of sight (l.o.s). Relation between $r$ and $s$ is given by
\bea
r = (s^2 + r_\odot^2 - 2sr_\odot \cos l \cos b)^{\frac{1}{2}},
\eea
where $r_\odot=8.5$ Kpc, the distance of the sun from the galactic centre
and $\rho_\odot=0.4$ GeV/cm$^3$ is the dark matter halo density at the
position of the solar system.   
The expression of the annihilation cross section ${\langle\sigma v \rangle}_{ss\rightarrow
\gamma \gamma}$ (in Eq. 22) for the process shown in Fig. \ref{gamma} is given in Ref.
\cite{gammacross}. 
In this calculation we have taken three different dark matter halo profiles (available in literature)
namely the Einasto profile \cite{einasto}, the NFW profile \cite{nfw} and the
Isothermal profile \cite{isothermal}. These halo profiles give the functional dependence
of $\rho(r)$ with $r$.
\begin{table}
\begin{center}
\vskip 0.5cm
\begin{tabular} {|c|c|c|c|c|}
\hline
{\bf Mass of Scalar} & {\bf Coupling $\delta_2$} & {\bf Flux using} & 
{\bf Flux using} & {\bf Flux using}\\
{\bf Dark Matter} & &{\bf Einasto Profile}&{\bf NFW Profile}&
{\bf Isothermal Profile}\\
GeV & & GeV cm$^{-2}$ s$^{-1}$ sr$^{-1}$ & GeV cm$^{-2}$ s$^{-1}$ sr$^{-1}$ &
GeV cm$^{-2}$ s$^{-1}$ sr$^{-1}$\\
\hline
130 &0.06&$1.971\times10^{-7}$&9.801$\times 10^{-8}$& $4.048\times 10^{-9}$\\
\hline
\end{tabular}
\end{center}
\caption{$\gamma-$ray flux obtained from the annihilation channel $SS\rightarrow
\gamma\gamma$, for three different dark matter halo profiles.}
\end{table} 
In the present calculation we have considered the value of the coupling $\delta_2 = 0.06$, which is
allowed by WMAP and all recent ongoing dark matter direct detection experiments that have been
considered in this work (Fig. \ref{constrain}, section 5).
We have calculated the $\gamma-$ray fluxes for all the halo three profiles considered
above and the results are shown in Table 1. 
The annihilation cross section ${\langle\sigma v \rangle}_{ss\rightarrow
\gamma \gamma}$ is calculated to be $7.13\times 10^{-31}$ cm$^3$/s for $\delta_2 = 0.06$.
From Ref. \cite {gammaref2}, one sees that the $\gamma$-ray flux obtained from 
``central region" of our galaxy is in the range
$4.0\times10^{-5}$ to $7.5\times10^{-5}$
(in GeV cm$^{-2}$s$^{-1}$sr$^{-1}$) (95{\%} CL) with best
fit value, $5.6\times10^{-5}$ GeV cm$^{-2}$s$^{-1}$sr$^{-1}$. From Table 1 we see that
in order to compare our results to those in Ref. \cite{gammaref2} the annihilation cross section for the
channel $SS\rightarrow \gamma \gamma$ in the present calculation must be enhanced by a factor of 
$\sim 3.0\times 10^2$ (for the Einasto profile) and $\sim 5.7\times 10^2$
(for the NFW profile) with respect to best fit value. As a result we
have to increase coupling $\delta_2$ from 0.06 to 1.03
(for the Einasto profile) and 1.43(for the NFW profile) respectively.
From Fig. {\ref {constrain}} it is seen that such a high value of $\delta_2$
is not satisfied by either any direct detection experiments
we have considered in this work or by the WMAP limits.
Therefore we conclude that a 130 GeV
dark matter in the present model can not explain the Fermi-LAT observed 130 GeV $\gamma-$ray line in the direction of the galactic centre unless
the process is boosted (by introducing a boost factor \cite{boost}) either by astrophysical justifications and/or by other particle physics methods.
\section {Summary and Conclusion}
In the present work we consider a simplest extension of SM by 
introducing a real gauge singlet (singlet under SU(2)$_{\rm L}
\times$U(1)$_{\rm Y}$) scalar $S$ to SM which can only interact 
with SM particles via Higgs. For the stability 
of $S$, $Z_2$ symmetry is imposed in the theory. 
Thus $S$ can be a viable candidate for cold dark matter. 
The scalar mass $m_S$ and the coupling are the two 
parameters in the theory. We have calculated the freeze out 
temperature  and relic density of this scalar dark matter candidate $S$ by 
solving Boltzmann's equation and have constrained the $m_S - \delta_2$ 
parameter space by using WMAP limit on relic density of dark matter in the 
universe and the results of recent ongoing dark matter direct search 
experiments like CDMS-II, DAMA, CoGeNT, XENON-10, XENON-100. We find that if 
$S$ is a dark matter candidate then its mass appears to be constrained  within 
two regions. One is a lower mass region where $m_S$ can vary from 6 GeV to 16 
GeV with $\delta_2$ lies in the limit $0.7 \leq \delta_2 \leq 1.25$
for $m_h= 120$ GeV. This region is supported by 
WMAP, CoGeNT and DAMA data. The other region is  higher mass region 
with the ranges for $m_S$ (in GeV) and $\delta_2$ found out 
to be $52.5 \la m_S \la 1000$, $0.02 \leq \delta_2 \leq 0.4$ 
for the same Higgs mass. This region is also supported by the
limits given by WMAP, CDMS-II, XENON-10, XENON-100, EDELWEISS-II experiments. 
We have calculated the possible differential direct 
detection rates and annual variations of 
total detection rates for scalar dark matter $S$ in case of 
two detector materials namely Ge, Xe. For all these target
materials we have found that differential detection rates decrease 
rapidly with the increase of observed recoil energy and they become 
vanishingly small for recoil energies beyond 10 keV for Ge 
with scalar mass $m_S = 10$ GeV. Whereas for Xe, the rates 
become vanishingly small for recoil energies beyond 80 keV
when $m_S = 55$ GeV. We have also shown how the total rates 
vary over a year for these target materials. These annual variations 
of total detection rates, if found, will be one sure evidence for 
dark matter detection. Finally in the last section we have calculated the
$\gamma-$ray flux for a 130 GeV scalar dark matter $S$ and we have
found that it is not possible to explain the Fermi-LAT observed
excess $\gamma-$ray line by a 130 GeV scalar dark matter, unless
a boost factor of order of $10^2$ is introduced with the annihilation
cross section of $SS\rightarrow\gamma\gamma$ channel.  

{\bf Acknowledgments:} A.B. thanks Debabrata Adak for some valuable 
discussions.  

\hspace{5cm}
\begin{center}
-----------------------
\end{center}
\end{document}